\documentclass[a4paper,11pt]{article}
\usepackage{jcappub} % for details on the use of the package, please see the JINST-author-manual
\usepackage{lineno}
\usepackage{mathtools}
\usepackage{macros}
\usepackage{comment}
\usepackage{amssymb,amsmath}
\usepackage{macros}
\usepackage{bm}
\usepackage{subfig}
\usepackage{color}
\usepackage{amsthm}
\usepackage{hyperref}
\usepackage{float}
\usepackage{tabularx}
\usepackage[normalem]{ulem}
\usepackage[utf8]{inputenc}
\usepackage[T1]{fontenc}
\usepackage{soul}

\usepackage{footnote}
\begin{document}

\title{ Magnetogenesis from axion-SU(2) inflation}
\date{}

\author[a,b,c,d]{Axel~Brandenburg,}
 \affiliation[a]{Nordita, KTH Royal Institute of Technology and Stockholm University, Hannes Alfv\'ens v\"ag 12, 106 91 Stockholm, Sweden}
 \author[a,b]{Oksana~Iarygina,}
 \author[e]{Evangelos~I.~Sfakianakis,}
 \author[f,a]{Ramkishor~Sharma}
 \affiliation[b]{The Oskar Klein Centre, Department of Astronomy, Stockholm University, 106 91 Stockholm, Sweden}
\affiliation[c]{McWilliams Center for Cosmology \& Department of Physics, Carnegie Mellon University, Pittsburgh, PA 15213, USA}
\affiliation[d]{School of Natural Sciences and Medicine, Ilia State University, 3-5 Cholokashvili Avenue, 0194 Tbilisi, Georgia}
\affiliation[e]{Department of Physics, Case Western Reserve University,
10900 Euclid Avenue, Cleveland, OH 44106, USA}
\affiliation[f]{CEICO, FZU-Institute of Physics of the Czech Academy of Sciences,
Na Slovance 1999/2, 182 00 Prague 8, Czech Republic}

% E-mail addresses: only for the corresponding author
\emailAdd{brandenb@nordita.org}
\emailAdd{oksana.iarygina@su.se}
\emailAdd{esfakianakis@ifae.es}
\emailAdd{sharma@fzu.cz}

\abstract
{We describe a novel proposal for inflationary magnetogenesis by identifying the non-Abelian sector of Spectator Chromo Natural Inflation (SCNI) with the $\rm{SU(2)}_{\rm L}$ sector of the Standard Model.
This mechanism relies on the recently discovered attractor of SCNI in the strong backreaction regime, where the gauge fields do not decay on super-horizon scales and their backreaction leads to a stable new trajectory for the rolling axion field. The large super-horizon gauge fields are partly transformed after the
electroweak phase transition into electromagnetic fields.
The strength and correlation length of the resulting helical magnetic fields depend on the inflationary Hubble scale and the details of the SCNI sector.
For suitable parameter choices we show that the strength of the resulting
magnetic fields having correlation lengths around $1~\rm Mpc$ are consistent with the required intergalactic magnetic
fields for explaining the spectra of high energy $\gamma$ rays from
distant blazars.}

\maketitle

\section{Introduction}

The presence of magnetic fields is ubiquitous in our universe \cite{Widrow:2002ud,Bernet:2008qp,Beck:2008ty,Kronberg:2007dy,Clarke_2001,Govoni:2004as,2005A&A...434...67V}.
Particularly intriguing is the evidence for extragalactic magnetic fields arising from the observations of distant blazars. The non-detection of the secondary GeV photons in blazar observations points towards the existence of extragalactic magnetic fields (EGMFs) between us (the observer) and the distant blazars \cite{Neronov:1900zz, Tavecchio:2010mk, Dolag:2010ni, Essey:2010nd, Taylor:2011bn, Takahashi:2013lba, Finke:2013tyq, Finke:2015ona}.
The strength of the magnetic field that is necessary to explain the blazar observations depends on its correlation length, $L$.
For $L\ge0.1~\rm Mpc$, the typical magnetic field $B_{0.1~\rm Mpc}$,
needs to be larger than $10^{-15}~\rm G$, or $10^{-17}~\rm G$,
depending on the assumptions made about the dynamics of the electromagnetic cascade and secondary GeV $\gamma$-ray emission \cite{Taylor:2011bn}.
For $L<0.1~\rm Mpc$, the typical magnetic field $B = B_{0.1~\rm Mpc} \sqrt{0.1~\rm Mpc/L}$ can be even larger.

While there is no conclusive answer to the question of the origin of EGMFs, many proposals have been put forth
(see, e.g., refs.~\cite{Kandus:2010nw,Widrow:2011hs,Durrer:2013pga,Subramanian:2015lua,Vachaspati:2020blt}).
For simplicity, we can categorize these proposals into those involving inflation and reheating \cite{Turner:1987bw, Ratra:1991bn,Carroll:1991zs,DiazGil:2007dy,Martin:2007ue,Demozzi:2009fu,Kanno:2009ei,Bamba:2006ga,Ng:2015ewp,Ferreira:2013sqa,Ferreira:2014hma,Campanelli:2013mea,Adshead:2016iae,Sharma:2017eps,Fujita:2019pmi,Gorbar:2021zlr,Kushwaha:2020nfa,Maity:2021qps,Tripathy:2021sfb}, and those focusing on early universe phase transitions (including QCD and electroweak) \cite{Vachaspati:1991nm, Sigl:1996dm, Dolgov:2001nv, Stevens:2007ep, Kahniashvili:2009qi, Henley:2010ba,Grasso:2000wj,Cheng:1994yr,Balaji:2024rvo}. 
In this work we focus on inflationary magnetogenesis, which encompasses a plethora of models and
continues to be a rich area of research.
 Since Maxwell's action is conformally invariant, there can be no significant magnetic field production during inflation, unless some way of breaking conformal invariance is introduced.
A simple way is to couple a U(1) gauge field, e.g.,
the electromagnetic (EM) or hypercharge gauge boson
of the Standard Model, to the rolling inflaton or some other dynamical (pseudo)scalar field during inflation. Usual couplings include the terms
$I(\phi) F_{\mu\nu} F^{\mu\nu} $ or $I(\phi) F_{\mu\nu} \tilde F^{\mu\nu}$, where $I(\phi)$ is some function of the scalar field $\phi$. The former is usually referred to as the Ratra model \cite{Ratra:1991bn} and we will not discuss it further (see, e.g., ref.~\cite{Barnaby:2012tk,Demozzi:2009fu} for non-Gaussianities and the strong coupling problem in the Ratra model).

We  focus on the axial coupling term $I(\phi) F_{\mu\nu} \tilde F^{\mu\nu}$, which results in the production of helical fields \cite{Campanelli:2008kh}. 
The importance of helical magnetic fields in the context of inflationary magnetogenesis has been mainly
based on the inverse cascade effect \cite{Campanelli:2007tc,Banerjee:2004df}.
During the inverse cascade process, power is transferred from short- to long-wavelength modes,
thereby slowing down the decay, and at the same time
increasing the correlation length \cite{Son:1998my}.
The coupling of axions to gauge fields during inflation has been extensively studied. The phenomenology includes  the amplification of parity violating gauge fields during slow-roll inflation \cite{Carroll:1991zs, Garretson:1992vt, Ferreira:2014zia} and their influence on the inflationary dynamics \cite{Prokopec:2001nc, Anber:2009ua, Barnaby:2011vw, Barnaby:2011qe}, as well as the generation of
metric fluctuations by a rolling auxiliary pseudo-scalar field during inflation \cite{Shiraishi:2013kxa, Cook:2013xea,Ferreira:2015omg,Peloso:2016gqs, Durrer:2024ibi}.  
The produced gauge fields can back-react on the axion \cite{Cheng:2015oqa, Notari:2016npn, Sobol:2019xls, DallAgata:2019yrr, Domcke:2020zez, Peloso:2022ovc, Caravano:2022epk, Garcia-Bellido:2023ser, vonEckardstein:2023gwk,Figueroa:2023oxc,Sharma:2024nfu}, possibly leading to significant non-Gaussianity \cite{Barnaby:2010vf, Caravano:2024xsb}. The strength of the axion-gauge coupling must be constrained to keep  non-Gaussianity of the density fluctuations, chiral gravitational waves, and the production of primordial black holes  within observational limits \cite{Barnaby:2010vf, Barnaby:2011vw, Barnaby:2011qe, Linde:2012bt, Bugaev:2013fya}. The authors of ref.~\cite{Adshead:2016iae} showed that a simple model of axion inflation coupled to the hypercharge field of the SM leads to very fast preheating; almost the entirety of the energy density of the inflaton is transferred to gauge fields within one $e$-fold after the end of inflation.
The resulting gauge field spectrum has a significant degree of helicity
and its amplitude is large enough to lead to present-day magnetic fields that are compatible with blazar observations. The parameters chosen allowed for both instantaneous preheating and efficient magnetogenesis, while at the same time not violating bounds from non-Gaussianity and primordial black hole production.

A seemingly straightforward extension of natural inflation \cite{Freese:1990rb} coupled to an Abelian gauge field, is the coupling of the axion-inflaton to a non-Abelian field instead. The non-trivial vacuum structure of the SU(2) sector \cite{ Maleknejad:2011jw, Maleknejad:2011sq} and its interplay with the axion field leads to a new inflationary attractor, in which the gauge field produces an extra source of friction, allowing for slow-roll inflation even in steep potentials \cite{Adshead:2013nka, Adshead:2013qp, Adshead:2012kp, Maleknejad:2016qjz,Papageorgiou:2018rfx}. This model goes under the name Chromo-Natural Inflation (CNI).
Similarly to the Abelian field case, the tensor modes of the SU(2) sector experience an instability, which causes one polarization to become exponentially amplified. The amplified SU(2) tensors seed gravitational waves, which are also chiral. The original version of Chromo-Natural inflation (one involving a cosine potential) has been shown to be 
incompatible with CMB observations \cite{Adshead:2013nka}, producing either too large tensor-to-scalar ratio $r$ or too small scalar spectral index $n_s$.
By invoking spontaneous breaking of the 
 SU(2) symmetry, the resulting model of Higgsed Chromo-Natural Inflation \cite{Adshead:2016omu} produces primordial observables which are observationally allowed
for certain parts of parameter space, while evading the Lyth bound \cite{Lyth:1996im} and generating observable gravitational waves at a lower inflationary scale. Furthermore, the resulting tensor spectral tilt $n_T$ generically violates the consistency relation $r=-8n_T$, where $r$ is the tensor to scalar ratio and $n_T$ is the spectral index of the tensor modes. Alternative ways to bring CNI in agreement with CMB data include modifying the potential of the axion field \cite{Maleknejad:2016qjz, Caldwell:2017chz},  delaying the CNI phase such that gravitational waves production happens at higher frequencies than CMB \cite{Obata:2014loa, Obata:2016tmo, Domcke:2018rvv} and  introducing non-minimal coupling to gravity \cite{Dimastrogiovanni:2023oid, Murata:2024urv}.
 Finally, by integrating out the axion field, a non-linear term is introduced involving the gauge field strength, which  leads to similar behavior and phenomenology \cite{Adshead:2012qe,Sheikh-Jabbari:2012tom, Maleknejad:2012dt, Maleknejad:2012fw, Iarygina:2021bxq}. 

An interesting extension of CNI was proposed in ref.~\cite{Dimastrogiovanni:2016fuu}, where the axion-SU(2) action was treated as a spectator sector. This allows the model to generate the tensor modes through the instability of the Chromo-Natural sector, while the scalar modes are produced by a dominant inflaton sector.
This decoupling of the inflationary energy scale from the gravitational wave (GW) amplitude allows for very low scale inflation with observable GWs~\cite{Fujita:2017jwq}. This model has been dubbed ``spectator chromo-natural inflation'' (SCNI) and has been shown to produce distinct GW spectra, depending on the shape of the axion potential \cite{Fujita:2018ndp}.

We recently explored the effects of backreaction on the SCNI model \cite{Iarygina:2023mtj},
where we reported the emergence of a novel attractor, supported by the backreaction of super-horizon gauge field modes on the rolling axion. This novel backreaction-supported attractor was found numerically and further described using analytical arguments. Our results were subsequently independently verified in ref.~\cite{Dimastrogiovanni:2024xvc}, where the possibility of primordial black hole formation was pointed out.
It is worth mentioning that pushing any nonlinear model to the strong backreaction regime can raise perturbativity issues. Ref.~\cite{Dimastrogiovanni:2024lzj} recently showed that perturbativity bounds on the parameter space are similar to those arising from the onset of the strong backreaction regime. That being said, computing perturbativity bounds inside the strong backreaction regime requires using the full numerical solution of the equations of motion. So as not to deviate from the main point of the current paper, that of inflationary magnetogenesis,  we leave the full analysis of perturbativity constraints for future work.

In the current work, we consider the model of spectator chromo-natural inflation, where we identify the SU(2) field as the weak sector of the Standard Model\footnote{A proposal to realize Chromo-Natural inflation using the Higgs as the inflaton \cite{Alexander:2023flr} may lead to similar phenomenology, but is beyond the scope of this work}. This can be thought of as the natural counterpart to natural inflation magnetogenesis, where the axion is coupled to the U(1) hypercharge sector. Even if the two models are similar in spirit, their analysis differs significantly due to the structure of the SCNI attractor and the richer structure of the fluctuations. The basic premise is the tachyonic generation of tensor modes of the SU(2) sector, equivalently weak bosons of the SM, during inflation. After the EW phase transition, the weak and hypercharge sectors mix and a component of the weak bosons is ``transformed'' into EM fields, which will also be helical to a large degree. After the generation of these EM fields, the electric part will be damped
by the primordial plasma, while the magnetic part will undergo inverse
cascading, thereby leading to the current length scales and field
strengths of magnetic fields across the universe today \cite{Banerjee:2004df,Brandenburg:2017neh}.
Furthermore, since we are interested in the evolution of the tensor modes from their generation during inflation
until the electroweak phase transition (EWPT), we revisited SCNI
by focusing on the end of inflation, which has been largely neglected in the literature so far. We thus point out the (rather generic) possibility of a second phase of inflation, where the chromo-natural sector dominates. In order to avoid that, one must either add extra couplings to drain the energy from the chromo-natural sector or adjust (tune) the initial value of the axion field, such that the spectator axion reaches its minimum before or at most shortly after the end of inflation.

An interesting point regarding inflationary magnetogenesis arises from the baryon isocurvature perturbations, which were computed in detail in ref.~\cite{Kamada:2020bmb}.
Due to the unknown details of the EWPT, the computation of the baryon number (and correspondingly the spatial variations of the baryon number) will necessarily include uncertainties (see, e.g., ref.~\cite{Kamada:2016cnb}). While ref.~\cite{Kamada:2020bmb} provides serious challenges on inflationary magnetogenesis, we leave a detailed evaluation of the baryon isocurvature perturbations in our model (and possible effects of BSM-generated alterations to the nature of the EWPT) for future work.

This work is organized as follows. In section~\ref{sec:model}, we review the spectator chromo-natural inflation model, its background dynamics and perturbations. Further we demonstrate when the SU(2) sector of the model is associated with the Standard Model weak bosons, the system inevitably enters the strong backreaction regime and converges to the recently discovered backreaction-supported attractor. In section~\ref{sec:Simulations} we investigate the dynamics of the background quantities and perturbations at the end of inflation.
We show that, when by the end of inflation the axion field does not reach the minimum of its potential,
the system enters a second inflationary phase dominated by the axion-SU(2) sector.
Further in section~\ref{sec:magnetogenesis}, we study the evolution of perturbations after inflation and discuss the magnetic field generation.
We conclude in section~\ref{sec:conclusions}.

\section{Model and attractor behavior}\label{sec:model}

In this section we introduce the model and background evolution of the system. We further discuss the backreaction constraints and indicate their immediate importance when the SU(2) sector of the model is associated with the Standard Model weak bosons. We provide a brief review of perturbations in axion-SU(2) inflation and their backreaction on the background evolution, along with subsequent convergence of axion field and the gauge field vacuum expectation value (VEV) to the new dynamical attractor.

\subsection{Background evolution}
The action for spectator axion-SU(2) inflation is given by \cite{Dimastrogiovanni:2016fuu}
\begin{equation}\label{eq:action}
S=\int d^4 x \sqrt{-g}\left( \frac{M_{\text{pl}}^2}{2}R -\frac{1}{2}(\partial \phi)^2-V(\phi)-\frac{1}{2}(\partial \chi)^2-U(\chi)
-\frac{1}{4}F^a_{\mu\nu}F^{a\, \mu\nu}+\frac{\lambda \chi}{4f}F^a_{\mu\nu}\tilde{F}^{a\, \mu\nu} \right),
\end{equation}
where $g\equiv\text{det} \, g_{\mu\nu}$, $R$ is the space-time Ricci scalar, $\phi(t)$ is the inflaton field,
$\chi(t)$ is the axion, $\lambda$ is the coupling constant between the gauge
and axion sectors, $f$ is the axion decay constant, and $M_\text{pl}$ is the reduced Planck mass.
The field strength of the SU(2) gauge field is
\begin{equation}
    F^a_{\mu\nu}=\partial_{\mu} A^a_{\nu}-\partial_{\nu} A^a_{\mu}-g \epsilon^{abc}A^b_{\mu}A^c_{\nu},
\end{equation}
with $g$ being the gauge field coupling. $\tilde{F}^{a\, \mu\nu}=\epsilon^{\mu\nu\rho\sigma}F^a_{\rho\sigma}/\left(2\sqrt{-\text{det} \, g_{\mu\nu}}\, \right)$
is a dual of the gauge field strength and $\epsilon^{\mu \nu\alpha\beta}$ is the antisymmetric tensor normalized as $\epsilon^{0123}=1$.

We use the axion potential of the form
\begin{equation}
    U(\chi)=\mu^4\left(1+\cos \frac{\chi}{f} \right),
\end{equation}
where $\mu$ is a constant that sets the energy scale of the axion. Without loss of generality, we restrict the axion field to be in the interval $\chi \in \left[0, \pi f \right]$.
In the SCNI model, the inflationary sector is responsible for the generation of the observed density fluctuations. Instead of modelling the inflationary dynamics as a quasi de-Sitter expansion, we choose to impose  concrete inflationary potentials, $V(\phi)$, which are specified in section \ref{sec:Simulations}. 

We work with the Friedmann-Robertson-Lema\^{i}tre–Walker (FRLW) metric
\begin{equation}
ds^2=-dt^2+a(t)^2 \delta_{ij}dx^i dx^j,
\end{equation}
where $i,j$ represent the spatial indices and $a(t)$ is the scale factor. For SCNI
the isotropic gauge field configuration at the background level is an attractor \cite{Maleknejad:2013npa} and is given by  \cite{Maleknejad:2011jw, Maleknejad:2011sq} 
\begin{gather}
A^a_0=0,\qquad
A^a_i=\delta ^a_i a(t)Q(t). \label{eq:isotropicQ}
\end{gather}

For the action \eqref{eq:action} and the isotropic gauge field configuration \eqref{eq:isotropicQ} the background system of equations for the inflaton field, axion and the gauge field vacuum expectation value is given by
\begin{gather}
M_{\text{pl}}^2\dot{H}=-\frac{1}{2}\dot{\phi}^2-\frac{1}{2}\dot{\chi}^2-\left( ( \dot{Q}+H Q)^2+ g^2 Q^4\right), \label{eq:back1} \\
3 M_{\text{pl}}^2 H^2=\frac{\dot{\phi}^2}{2}+V(\phi)+\frac{\dot{\chi}^2}{2}+U(\chi)+\frac{3}{2}\left(\dot{Q}+H Q\right)^2+\frac{3}{2}g^2 Q^4, \label{eq:back2}\\
 \ddot{Q}+3H\dot{Q}+\left(\dot{H}+2H^2\right)Q+2g^2Q^3=\frac{g\lambda }{f}\dot{\chi}Q^2,\\
 \ddot{\chi}+3H\dot{\chi}+U_{\chi}(\chi)=-\frac{3g\lambda}{f}Q^2\left( \dot{Q}+HQ\right),\\
 \ddot{\phi}+3H\dot{\phi}+V_{\phi}(\phi)=0, \label{eq:back3}
\end{gather}
where an overdot denotes a derivative with respect to cosmic time $t$, $H=\dot{a}/a$ is the Hubble parameter, $V_{\phi}(\phi)=\partial V(\phi)/\partial \phi$ and $U_{\chi}(\chi)=\partial U(\chi)/\partial \chi$.

A viable inflationary model requires $f,\mu \ll M_{\rm pl}$ with the energy scale of
inflation being well below the cut-off of the effective theory
$ {f}/{\lambda} \gg H$. Furthermore, the existence of the chromo-natural attractor solution restricts \cite{Adshead:2012kp} the parameter space to
  $\Lambda \equiv \lambda Q/f \gg {\rm min}(\sqrt{2} , \sqrt{3}H/gQ)$.  When the above conditions are satisfied, the chromo-natural inflation model in the slow-roll approximation approaches an attractor  \cite{Adshead:2012kp, Adshead:2013nka}
\begin{equation}
    \begin{split}
 \frac{\lambda}{f}\dot{\chi}=2gQ+\frac{2H^2}{gQ},\qquad
 \dot{Q}=-HQ+\frac{f}{3g\lambda}\frac{U_{\chi}}{Q^2}. \label{eq:attractorCNI}
\end{split}
\end{equation}

Moreover, to ensure that scalar perturbations are controlled by the
inflaton field, we impose the spectator condition that the energy
densities of the axion field and gauge sector are subdominant to
that of the inflaton, i.e.,
\begin{equation}
    \rho_{\phi}\gg \rho_{Q_{E}},\,\rho_{Q_B}, \, \rho_{\chi}.
\end{equation}
where the corresponding energy densities are given by
\begin{eqnarray}\label{eq:rho}
\rho_{\phi}=\frac{1}{2}\dot{\phi}^2+V(\phi), \quad
\rho_{Q_E}=\frac{3}{2}( \dot{Q}+H Q)^2,\quad
\rho_{Q_B}=\frac{3}{2}g^2 Q^4, \quad
\rho_{\chi}=\frac{1}{2}\dot{\chi}^2+U(\chi).
\end{eqnarray}
The stability of scalar perturbations of the gauge sector requires $ g Q/H >\sqrt{2}$. We impose these criteria to be satisfied for the part of inflation that corresponds to CMB scales\footnote{In this work, we focus on the study of tensor perturbations and leave a detailed analysis of scalar perturbations for future work. }. 

Finally, we assume that the inflaton field $\phi$, along with the spectator sector comprised of the axion $\chi$ and the VEV of the gauge field $Q$, are slowly rolling during inflation, i.e., their slow-roll parameters are smaller than unity. The first slow-roll parameter $\epsilon_H$ is defined as
\begin{equation}
    \epsilon_H=-\frac{\dot{H}}{H^2}=\epsilon_{\phi}+\epsilon_{Q_E}+\epsilon_{Q_B}+\epsilon_{\chi}, \label{eq:epsilonH}
\end{equation}
with the contributions 
\begin{gather}
\epsilon_{\phi}=\frac{\dot{\phi}^2}{2M_{\text{pl}}^2 H^2},\quad \epsilon_{Q_E}=\frac{( \dot{Q}+H Q)^2}{M_{\text{pl}}^2 H^2},\quad \epsilon_{Q_B}=\frac{g^2Q^4}{M_{\text{pl}}^2 H^2},\quad
\epsilon_{\chi}=\frac{\dot{\chi}^2}{2M_{\text{pl}}^2 H^2}.
\end{gather}

\subsection{Standard Model axion-SU(2) sector and backreaction}
To study magnetogenesis from axion-SU(2) dynamics during inflation, we
associate the Standard Model weak bosons with the SU(2) sector of the model\footnote{The axion coupled to the SM weak bosons has a mass of $m_\chi = \mu^2/f\gtrsim 10^{11} \, {\rm {GeV}}$ (see table~\ref{table1}), which makes its effects in accelerator experiments unobservably small.}.
Using the SM restricts the gauge coupling to be $g={\cal O}(0.1)$, where we leave room for possible changes of the renormalization group flow to large energies due to unknown physics. As we will see, large gauge field couplings immediately push the system towards the strong backreaction regime. 

It is convenient to introduce the parameters $m_Q$ and $\xi$,
\begin{equation}\label{mQxidef}
    m_Q=\frac{g Q}{H}, \qquad \xi=\frac{\lambda}{2fH}\dot{\chi},
\end{equation}
which control the amplification of the gauge field 
fluctuations around horizon crossing and the subsequent sourcing of gravitational waves. When $m_Q$ is approximately constant, the backreaction is small and can be neglected when \cite{Papageorgiou:2019ecb}
\begin{equation}
    g \ll \left( \frac{24\pi^2}{2.3\cdot e^{3.9 m_Q}}\frac{1}{1+m_Q^{-2}}\right)^{1/2}. \label{eq:constraintBR}
\end{equation}
Taking the smallest allowed value of $m_Q=\sqrt{2}$ to ensure the stability of scalar perturbations and evaluating the right-hand side of eq.~\eqref{eq:constraintBR}
we get $g\ll 0.53$.
It thus follows that gauge-field couplings  $g\gtrsim {\cal O}(0.1)$ inevitably push the system into the strong backreaction regime.

\subsection{Perturbations and the backreaction-supported attractor }

The backreaction in chromo-natural inflation is caused by the tachyonic amplification of  gauge field modes, which in turn backreact on the background dynamics and change it. 

We choose the gauge and decomposition for field fluctuations following the ref.~\cite{Dimastrogiovanni:2012ew}
\begin{equation}\label{eq:decomposition}
    \begin{split}
     \phi &= \phi+\delta\phi,\\
        \chi &= \chi+\delta\chi,\\
       W_{0}^{a}&=a (Y_{a}+\partial_{a}Y),\\
W_{i}^{a}&=a\left[\left(Q+\delta Q\right)\delta_{ai}+\partial_{i}\left(M_{a}+\partial_{a}M\right)+\epsilon_{iac}\left(U_{c}+\partial_{c}U\right)+t_{ia}\right]\,,\\
g_{00}&=-a^2\left(1-2\varphi \right),\\
g_{0i}&=a^2\left(B_i+\partial_i B\right),\\
g_{ij}&=a^2\left[\left(1+2\psi \right)\delta_{ij}+2\partial_i\partial_j E+\partial_i E_j+\partial_j E_i+h_{ij} \right],
    \end{split}
\end{equation}
where $a=1,2,3$ is the SU(2) index (not to be confused with the scale factor $a(t)$) and $i=1,2,3$ is the index for spacial coordinates.
Furthermore, $t_{ia}$ and $h_{ij}$ are the transverse and traceless tensor modes of gauge field and metric respectively.
Transverse vector modes are  $Y_a, M_a, U_c, B_i, E_i$ and scalar modes $\delta \phi,\, \delta \chi,\, Y,\, \delta Q,\, M,\, \varphi,\, B $\,\footnote{The scalar perturbation should not be confused with the magnetic field strength $B$.}. 

We focus on the tensor modes of the gauge field $t_{ia}\subset \delta W_{i}^{a}$ and the metric $h_{ij}\subset \delta g_{ij}$ and expand them  into a helicity basis in Fourier space:
\begin{equation}\label{eq:tijhij}
    \begin{split}
 t_{ij}(\bold{x}, \tau)=\int \frac{d^3 k}{(2 \pi)^{3/2}}e^{i \vec{k}\cdot\vec{x}}\sum_{\nu=\pm}   \Pi^*_{ij, \nu}(\vec{k})t_{\nu}(\tau, \vec{k}),\\
 h_{ij}(\bold{x}, \tau)=\int \frac{d^3 k}{(2 \pi)^{3/2}}e^{i \vec{k}\cdot\vec{x}}\sum_{\nu=\pm}   \Pi^*_{ij, \nu}(\vec{k})h_{\nu}(\tau, \vec{k}),
   \end{split}
\end{equation}
    where the polarization tensors satisfy \cite{Namba:2015gja}
    \begin{equation}
        \Pi^*_{ij, \pm}(\vec{k})=\epsilon_{i, \pm}(\vec{k})\epsilon_{j, \pm}(\vec{k}),
    \end{equation}
        and $\epsilon_{i,\pm}$ are helicity vectors with the properties
\begin{equation}
    \begin{split}
            \vec{k}\cdot \vec{\epsilon}_{\pm}(\vec{k})=0,& \quad \vec{k}\times \vec{\epsilon}_{\pm}(\vec{k})= \mp i k \vec{\epsilon}_{\pm}(\vec{k}),\\
            \vec{\epsilon}_{\nu}(\vec{k})\cdot \vec{\epsilon}\, {}^*_{\nu'}(\vec{k})=\delta_{\nu \nu'},& \quad  \vec{\epsilon}\, {}^{*}_\pm(\vec{k})= \vec{\epsilon}_{\pm}(-\vec{k})= \vec{\epsilon}_{\mp}(\vec{k}).
   \end{split}
\end{equation}
        We denote right- and left-handed modes $(\pm)$ as $(R,L)$ and 
  canonically normalize perturbations as
\begin{equation}
h_{R,L}=\frac{\sqrt{2}}{M_{\rm pl} a}\psi_{R,L},\qquad t_{L,R}=\frac{1}{\sqrt{2}a}T_{R,L}. \label{eq:thnormalised}
\end{equation}
The equations of motion for perturbations up to ${\cal O}(\epsilon)$ are given by 
\begin{gather}
        \pp_{t}\psi_{R,L}+H \p_t \psi_{R,L}+ \left(\frac{k^2}{a^2}-2 H^2\right)\psi_{R,L}=-2 H \sqrt{\epsilon_{Q_E}}\p_{t}T_{R,L}+2 H^2 \sqrt{\epsilon_{Q_B}}\left(m_Q\mp \frac{k}{a H}\right)T_{R,L},\label{eq:perturbPsi}\\
        \pp_{t}T_{R,L}+H \p_t T_{R,L}+\left(\frac{k^2}{a^2}+2 H^2(m_Q \xi\mp \frac{k}{a H}(m_Q+\xi))\right)T_{R,L}=2 H \sqrt{\epsilon_{Q_E}}\p_t \psi_{R,L}\nonumber\\
        +2 H^2(\sqrt{\epsilon_{Q_B}}\left(m_Q\mp \frac{k}{a H}\right)
        +\sqrt{\epsilon_{Q_E}})\psi_{R,L}.\label{eq:perturbT}
\end{gather}

The exponential growth of the tensor modes backreacts on the background equations of motion \cite{Dimastrogiovanni:2016fuu, Fujita:2017jwq,  Maleknejad:2018nxz,  Papageorgiou:2019ecb, Ishiwata:2021yne}.
To take into account the contribution from backreaction, the background equations of motion can be written as 
\begin{gather}
\ddot{Q}+3H\dot{Q}+\left(\dot{H}+2H^2\right)Q+2g^2Q^3-\frac{g\lambda }{f}\dot{\chi}Q^2+{\cal T}^Q_{\rm BR}=0,\label{qeqn}\\
 \ddot{\chi}+3H\dot{\chi}+U_{\chi}(\chi)+\frac{3g\lambda}{f}Q^2\left( \dot{Q}+HQ\right)+{\cal T}^{\chi}_{\rm BR}=0.\label{chieqn}
\end{gather}
The backreaction terms $ {\cal T}^Q_{\rm BR}$ and ${\cal T}^{\chi}_{\rm BR}$ contain the integrals over the mode functions and are defined as\footnote{For homogeneous backreaction the effect from spatial gradients of inflation and axion fields are neglected. }
\begin{gather}
{\cal T}^Q_{\rm BR}=\frac{g}{3a^2}\int \frac{d^3 k}{(2\pi)^3}\left( \xi H -\frac{k}{a}\right)|T_R|^2,\label{qbackint}\\
 {\cal T}^{\chi}_{\rm BR}=-\frac{\lambda}{2 a^3 f}\frac{d}{dt}\int  \frac{d^3 k}{(2\pi)^3}\left(a\, m_Q H-k\right)|T_R|^2.\label{chibackint}
\end{gather}

The effect of homogeneous backreaction during axion-SU(2) dynamics was recently explored in ref.~\cite{Iarygina:2023mtj}. When the backreaction becomes strong, the solution converges to \textit{the new dynamical attractor} with  negative values of the gauge field VEV and  decreased velocity of the axion field, given by \cite{Iarygina:2023mtj}
\begin{gather}
    \frac{\lambda}{ f}\dot{\chi}\simeq -\frac{2 H^2}{g Q}, \quad
    U_{\chi}=-\frac{3 g\lambda}{f}HQ^3+\frac{1}{\tilde{\alpha}}\left(4H^2 Q+2g^2Q^3    \right), \label{eq:newattractor}
\end{gather}
where the parameter $\tilde{\alpha}$ is the ratio of the backreaction integrals
\begin{eqnarray}
\tilde{\alpha}=\frac{{\cal T}^{Q}_{\rm BR}}{{\cal T}^{\chi}_{\rm BR}}\simeq \frac{2}{9}\frac{Hf}{\lambda g \, Q^2} \label{eq:alphanewAtr}
\end{eqnarray}
and $Q$ is the VEV of the gauge field on the new attractor. The new solution \eqref{eq:newattractor} resembles the chromo-natural attractor solution \eqref{eq:attractorCNI}, with an opposite sign for the VEV $Q$, smaller axion velocity, and a modified dependence of $Q$ on $U_{\chi}$.

Ref.~\cite{Iarygina:2023mtj} approximated inflation as having a constant Hubble parameter. Therefore, once the axion-SU(2) system converged to the new backreaction-supported attractor, it remained there indefinitely.
In the current work we aim to trace the evolution of the gauge field perturbations from the end of inflation until the EWPT,
in order to investigate the generation and evolution of magnetic fields. In the next section, we go beyond the results of ref.~\cite{Iarygina:2023mtj}, by including the time dependence of the Hubble parameter in order to probe the dynamics of the gauge field perturbations and the background mode through the end of inflation.

\section{Gauge-axion dynamics until the end of inflation }\label{sec:Simulations}

In this section we study the generation and evolution of tensor perturbations and model their evolution through the end of inflation. To depart from 
the approximation of a constant Hubble rate during inflation, we ran several simulations for different inflationary background models, specifically using quadratic and $\alpha$-attractor potentials.
The $\alpha$-attractor potential is in agreement with CMB data and carries significant theoretical motivation \cite{Kallosh:2013hoa, Kallosh:2013yoa}, whereas the quadratic potential can be thought of as an approximation of a more complicated potential, valid near the end of inflation, where the inflaton behaves as a massive scalar field. 
We provide the details of the parameters used for the different inflationary background models in section \ref{sec:backgroundModels}. The simulation parameters are summarized in \Tab{table1}. The initial values of $\mu$ and $\chi/f$ were chosen to ensure that the initial Hubble parameter is the same and that the axion field $\chi$ approaches one of the minima of its potential at $f\pi/2$
before the end of inflation, except for run F. If $\chi$ does not reach the minimum of its potential before the end of inflation, the system will enter a second inflationary phase dominated by the axion-SU(2) sector, as we demonstrate later.
This has been largely neglected in the spectator CNI literature so far and provides an important constraint on the viable parameter space of these models.
\begin{table}
\begin{center}
\begin{tabular}{cm{0.4cm}cccccccr}
\hline
Run & $g$ & $\mu$ & $\chi_i/f$ & $Q_i$ & $m_{Q_i}$ & $H_i$ &$\!\!n_{\min}\!\!$ &$\!\!n_{\max}\!\!$ & model \\
\hline
A1 & $0.1$ & $5.35 \times 10^{-4}$ & $0.975 \pi $ & $1.26 \times 10^{-4}$& $2.37$ & $5.3 \times 10^{-6}$ & $1$ & $11$ & Const H \\ % mu5p35em4_g0p1_2048a_ct_constH
A2 & $0.1$ & $4.76 \times 10^{-4}$ & $0.96 \pi $ & $1.26 \times 10^{-4}$& $2.37$ & $5.3 \times 10^{-6}$ & $1$ & $11$ & Const H\\ % mu4p76em4_g0p1_2048a_ct_constH
B & $0.1$ & $5.80 \times 10^{-4}$ & $0.982 \pi $ & $1.26 \times 10^{-4}$ & $2.36$ & $5.3 \times 10^{-6}$ & $1$ & $8$ & Quadratic \\ % mu5p8em4_g0p1_2048a_ct_lhubble
C1 & $0.1$ & $5.35 \times 10^{-4}$ & $0.975 \pi $ & $1.26 \times 10^{-4}$ & $2.37$ & $5.3 \times 10^{-6}$ & $1$ & $11$ & $\alpha$-attractor \\ % mu5p35em4_g0p1_2048a_ct_lhubble_alpha_attract_rep
C2 & $0.1$ & $5.30 \times 10^{-4}$ & $0.974 \pi $ & $1.26 \times 10^{-4}$ & $2.37$ & $5.3 \times 10^{-6}$ & $1$ & $11$ & $\alpha$-attractor \\ % mu5p29em4_g0p1_2048a_ct_lhubble_alpha_attract
C3 & $0.1$ & $5.25 \times 10^{-4}$ & $0.973 \pi $ & $1.26 \times 10^{-4}$ & $2.37$ & $5.3 \times 10^{-6}$ & $1$ & $11$ & $\alpha$-attractor \\ % mu5p24em4_g0p1_2048a_ct_lhubble_alpha_attract
D & $0.65$ & $1.60 \times 10^{-4}$ & $0.975 \pi $ & $1.35 \times 10^{-5}$ & $1.65$ & $5.3 \times 10^{-6}$ & $-2$ & $8$ & $\alpha$-attractor \\ % mu1p6em4_g0p65_2048a_ct_lhubble_alpha_attract
E & $0.65$ & $1.57 \times 10^{-4}$ & $0.973 \pi $ & $1.35 \times 10^{-5}$ & $1.65$ & $5.3 \times 10^{-6}$ & $-2$ & $8$ & $\alpha$-attractor \\ % mu1p57em4_g0p65_2048a_ct_lhubble_alpha_attract
F & $0.1$ & $3.79 \times 10^{-4}$ & $0.9 \pi $ & $1.26 \times 10^{-4}$ & $2.37$ & $5.3 \times 10^{-6}$ & $1$ & $8$ & $\alpha$-attractor \\ % mu3p79em4_g0p1_2048a_ct_lhubble_alpha_attract
G & $0.1$ & $1.24 \times 10^{-3}$ & $0.988 \pi $ & $3.04 \times 10^{-4}$ & $5.71$ & $5.3 \times 10^{-6}$ & $1$ & $8$ & $\alpha$-attractor \\ % mu1p242em3_g0p1_2048a_ct_lhubble_alpha_attract
H & $0.1$ & $5.25 \times 10^{-4}$ & $0.974 \pi $ & $1.25 \times 10^{-4}$ & $2.34$ & $5.3 \times 10^{-6}$ & $1$ & $11$ & $\alpha$-attractor \\ % mu5p25em4_g0p1_2048a_ct_lhubble_alpha_attract_rep2
\hline
\end{tabular}
\caption{Parameters for the runs discussed in the paper. For all these runs $\lambda=2000$. The value of $f=0.09$ for runs with $g=0.1$ and $f=0.009$ for runs with $g=0.65$. Here $H_i$ represents the value of Hubble parameter at the start of the simulations.  All dimensionful quantities are measured in units of $M_{\rm{Pl}}$.}\label{table1}
\end{center}
\end{table}

\subsection{Background inflaton models}\label{sec:backgroundModels}

To probe the end of inflation, we use quadratic and $\alpha$-attractor inflationary models for the background evolution. The details of the models are provided below.
For the quadratic model, the potential of the scalar field has the usual form
\begin{equation}
    V_{\rm{quad}}=\frac{1}{2}m^2\phi^2.
\end{equation}
To achieve roughly sixty $e$-folds of inflation, we choose $\phi_i=15.56\, M_{\rm {Pl}}$. The relevant
mass scale is chosen as $m=8.4 \times 10^{-7}\, M_{\rm {Pl}}$. These values lead to $H_i=5.33 \times 10^{-6}\, M_{\rm {Pl}}$ at the initial time, which we take to coincide with the beginning of inflation.

In the case of $\alpha$-attractors, the potential for the scalar field is given by,
\begin{equation}
    V_\textrm{$\alpha$-attr}(\phi)=\alpha M \left(\left(\tanh{\left(\beta\phi/2 \right)}\right)^2 \right)^n \, ,
\end{equation}
where $\beta= \sqrt{2/3\alpha}$. In our simulations, we chose $\alpha=1, \,  M=8.7 \times 10^{-11}\, M_{\rm {Pl}}, \, n=3/2$, and the initial value of the scalar field, $\phi_{\rm in}=6.7\, M_{\rm {Pl}}$, to achieve around sixty $e$-folds of inflation. In this case, $H_i=5.32 \times 10^{-6}\, M_{\rm {Pl}}$ at the beginning of inflation.

\subsection{Dynamics during inflation and second inflationary phase }
To realize the SU(2) sector as the Standard Model SU$_{\rm L}$(2) sector, we consider gauge field couplings $g={\cal O}(0.1)$. In our simulation, we examine two cases: $g=0.1$ (runs A1, A2, B, and C1-3) and $g=0.65$ (runs D and E). The backreaction bound given in \Eq{eq:constraintBR} suggests that the value of $m_Q$ should be less than 1.3 for the $g=0.65$ case and less than 2.4 for the $g=0.1$ case to avoid backreaction of the tensor perturbations of the SU(2) gauge fields on its background evolution. It is important to note that, even if the value of $m_Q$ is below the backreaction bound initially, backreaction may still become important at a later epoch during inflation (see the run $\mu3$ in ref.~\cite{Iarygina:2023mtj}). We also consider $m_Q> \sqrt{2}$ to avoid instability of the scalar perturbations of the SU(2) sector. Therefore, for the case of $g=0.65$, backreaction will be significant from the beginning since $m_Q > \sqrt{2}$ already lies within the backreaction regime. However, for the case of $g=0.1$, by properly choosing a value of $m_Q$ smaller than 2.6, backreaction will not be important initially, but can become significant at a later stage in the evolution.

For the numerical simulations, we use the {\sc Pencil Code} \cite{JOSS} and solve the background equations \eqref{eq:back1}, \eqref{eq:back2}, \eqref{eq:back3}, \eqref{qeqn}--\eqref{chibackint} with perturbation equations \eqref{eq:perturbPsi}--\eqref{eq:perturbT}. The simulations are performed in cosmic time. Similarly as in ref.~\cite{Iarygina:2023mtj}, we set the initial conditions for the real and imaginary parts of the perturbation variables as
\begin{align}
    T_{R,L}=\frac{1}{\sqrt{2k}}e^{ik/a_i H_i}, \quad    \p_t  T_{R,L}=-\frac{i}{a_i}\sqrt{\frac{k}{2}}e^{ik/a_i H_i},
\end{align}
(similarly for $\psi_{R,L}$) with $a_i=1$. The contributions from quantum vacuum fluctuations of $T_{R,L}$ in the calculation of the backreaction integrals in eqs.~\eqs{qbackint}{chibackint} are discarded by setting $|T_{R,L}|^2$ to zero
when $|T_{R,L}|^2<1/2k$.
In our simulations, we have $n_k$ points of $k$ in the range
\begin{equation}
    n_{\min}\le \ln (k/a_i H_i) \le n_{\max}.
\end{equation}
The $n_k$ points are chosen such that they are distributed uniformly in $\ln k$ and $n_k=2048$ for all our simulations. The values of $n_{\min}$ and $n_{\max}$ are provided in \Tab{table1} for each run.

\begin{figure}[h!]
\centering
 \includegraphics[width=1\textwidth]{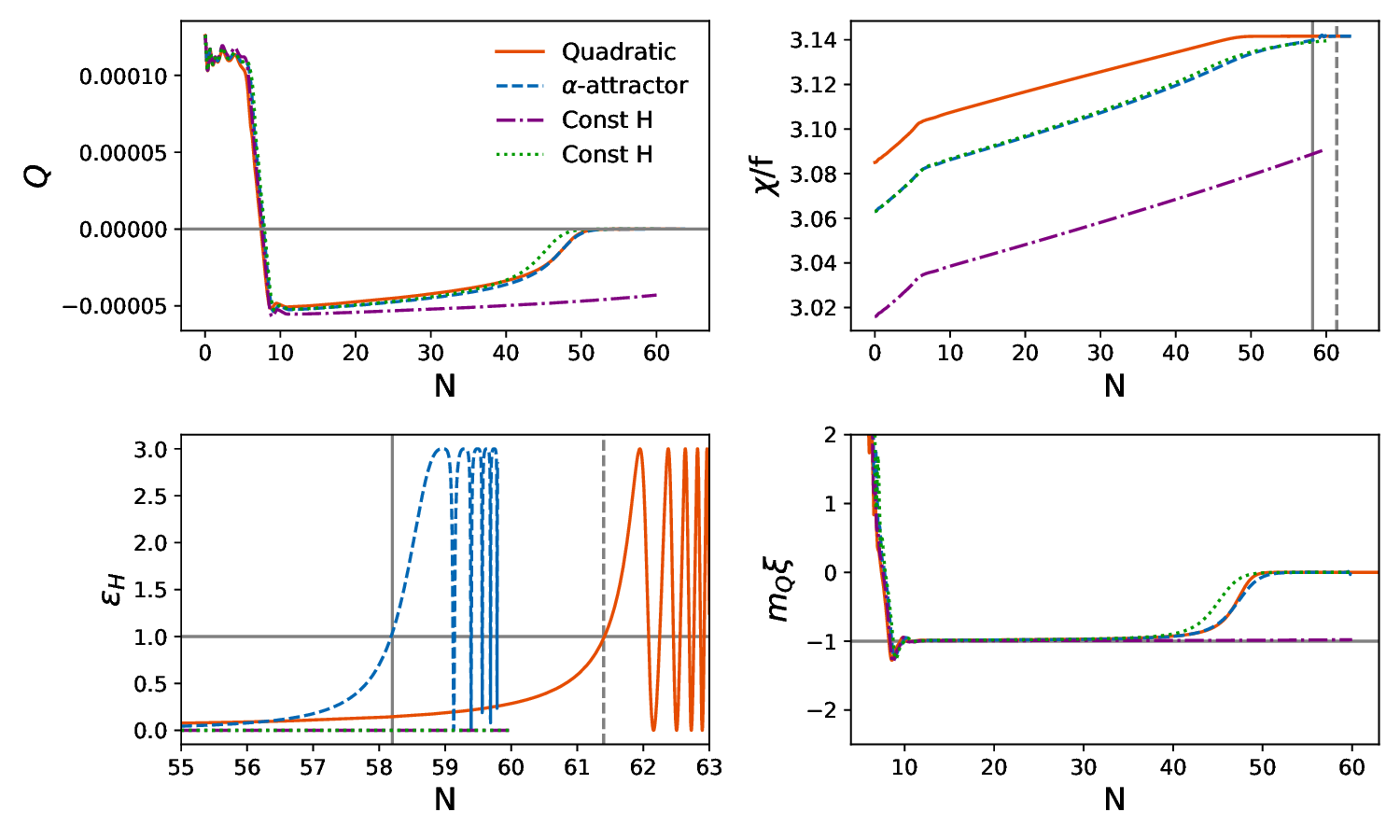}
\caption{The evolution of gauge field VEV $Q$ (top left), axion field $\chi/f$ (top right), the first slow-roll parameter $\epsilon_H$ (bottom left)  and the combination $m_Q \xi$ (bottom right) with respect to the number of $e$-folds $N$ for different simulations. The solid orange curve corresponds to (run B, quadratic inflation) and the dashed blue to (run C1,\, $\alpha$-attractors). The dot-dashed green (run A1, const H) and dot-dashed purple (run A2, const H) curves relate to scenarios where $H$ remains constant during inflation but with different initial values of $\chi/f$: smaller for run A2 and bigger for run A1. The solid and dashed grey grid lines correspond to the end of inflation for runs C1 and B, respectively. The color coding is the same for the whole panel. The parameters for each run are shown in table~\ref{table1}.  } 
\label{Q_and_chi_plot_mu2p69em4_2048a_ct_lhubble}
\end{figure}

We show the background evolution of the gauge field VEV $Q$ and axion field $\chi/f$  in the upper left and right panels of \Fig{Q_and_chi_plot_mu2p69em4_2048a_ct_lhubble}, respectively.
We set $(\pi - \chi/f)<0.1$ initially, so that the
axion and gauge field VEV relax to their respective minima (before or) close to the end of inflation.
The solid orange and dashed blue curves correspond to the quadratic and $\alpha$-attractor inflationary models, respectively. For $g=0.1$ and $m_Q\simeq 2.37$, the backreaction of the perturbations is significant, which forces $Q$ to settle into the negative attractor solution \eqref{eq:newattractor}--\eqref{eq:alphanewAtr} and reduces the velocity of the $\chi$ field, as discussed in refs.~\cite{Iarygina:2023mtj,Dimastrogiovanni:2024xvc}. As $\chi$ approaches a minimum of its potential, $Q$ tends to zero and remains there, transitioning from a chromo-natural attractor into the trivial vacuum of the theory. The transition to zero is occurring smoothly for all the runs considered; see \Fig{Q_and_chi_plot_mu2p69em4_2048a_ct_lhubble} with solid orange curve (run B, quadratic inflation), dashed blue (run C1,\, $\alpha$-attractors), dot-dashed green (run A1, const H), and dot-dashed purple (run A2, const H). For the ``const H'' runs, the Hubble parameter is constant during inflation, but we choose a different initial value for the axion field in each run; see table~\ref{table1}. 
The initial parameters are such that the end of inflation occurs at $N=61.4$ for the quadratic model and $N=58.2$ for the $\alpha$-attractor model. The end of inflation is defined as the time when the first slow-roll parameter  $\epsilon_H$ \eqref{eq:epsilonH} reaches unity. In the lower panels of \Fig{Q_and_chi_plot_mu2p69em4_2048a_ct_lhubble}, we show the time evolution of $\epsilon_H$ (bottom left panel) and $m_Q \xi$ (bottom right panel), defined by eq.~\eqref{mQxidef}. During inflation, when the Hubble parameter is approximately constant, we see that $m_Q \xi\simeq -1$ and the system stays at the  backreaction-supported attractor with $Q<0$; see appendix~D of ref.~\cite{Iarygina:2023mtj}. When the axion relaxes into a minimum of its potential, $m_Q \xi$ vanishes.

\begin{figure}[h!]
\centering
 \includegraphics[width=1\textwidth]{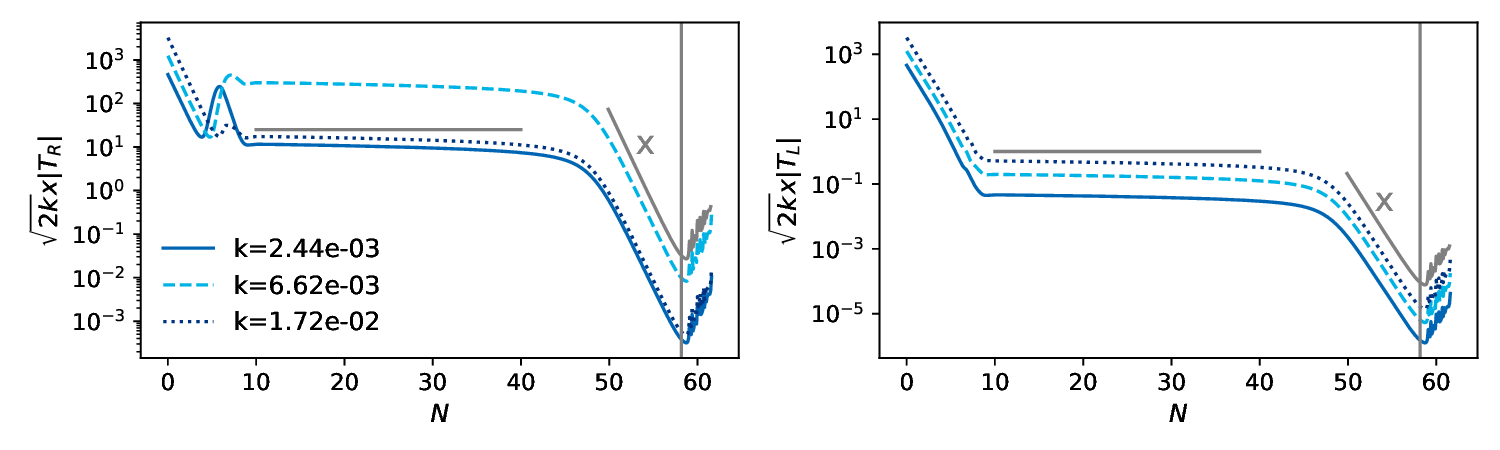}
\caption{The evolution of tensor perturbations of gauge field $T_{R,L}$ with respect to the number of $e$-folds $N$ for there different $k$-values for run (C1,\, $\alpha$-attractors). The grey vertical lines represent the end of inflation.
The two grey line segments designate a constant value and a function proportional to $x$. The values of $k$ shown in the legend are normalized by $M_{\rm pl}$.}
\label{tr_and_psi_plot}
\end{figure}

Tensor perturbations of the gauge field will eventually seed magnetic fields. Hence, it is crucial to investigate the dynamics of perturbations as $Q$ approaches zero. We show the evolution of gauge field perturbations $T_{R,L}$ in \Fig{tr_and_psi_plot} for three different values of the comoving wave number $k$. 
Before the transition of $Q\to 0$, the evolution of $T_R$ for the super-horizon modes is such that $\sqrt{2k}\,xT_R(x)$ remains (roughly) constant in time\footnote{The evolving Hubble scale near the end of inflation leads to a $m_Q\xi$ deviating slightly from $-1$ and thus $xT_R$ being almost constant but not exactly so.}, where $x\simeq k/(aH)$.
This is derived through the equation for $T_R$
\begin{equation}
    \pp_{x}T_{R,L}+\left(1+\frac{2}{x^2}m_Q \xi\right)T_{R,L}\simeq 0 \,; \label{eq:TRattractor}
\end{equation}
see appendix~D of ref.~\cite{Iarygina:2023mtj} for more details. At the backreaction-supported attractor $m_Q \xi \simeq -1$, and by considering the super-horizon regime where $x \simeq k/aH\ll 1$, we see that
 $T_{R/L} \propto 1/x$ and thus the combination $x  T_{R/L}$ remains constant as long as the system follows the backreaction-supported attractor. We can define the (almost) constant value of $x T_{R/L}$ 
 during the attractor through
 \begin{equation}
     \sqrt{2k}\, x T_R = c_1(k)
 \end{equation}
where the constant $c_1$ is different for each wave number $k$.

Before transitioning to the next section and discussing the evolution of perturbations around the end of inflation, we demonstrate the possibility of a second inflationary phase. If the $\chi$ field is initialized such that its initial value is far from the one corresponding to the minimum of its potential ($\chi/f=\pi$), the axion field will not reach to its minimum by the end of inflation and the total energy density of the $\chi$ field will remain dominated by its potential energy. Since the energy density of the inflaton field decreases after the end of inflation, the total energy density of the universe will eventually be dominated by the almost constant potential energy of the $\chi$ field. At this point, the system will enter a second inflationary era, dominated by the potential energy of the axion, similarly to chromo-natural inflation.\footnote{ Interestingly though, at least initially, the first slow-roll parameter $\epsilon_H$ is strongly affected by the oscillating field $\phi$ and thus exhibits itself oscillations on top of an average value of $\epsilon_H<1$. When these die out, we expect this second inflationary stage to be identical to  ``standard'' chromo-natural inflation.}
\begin{figure}[h!]
\centering
\includegraphics[width=1\textwidth]{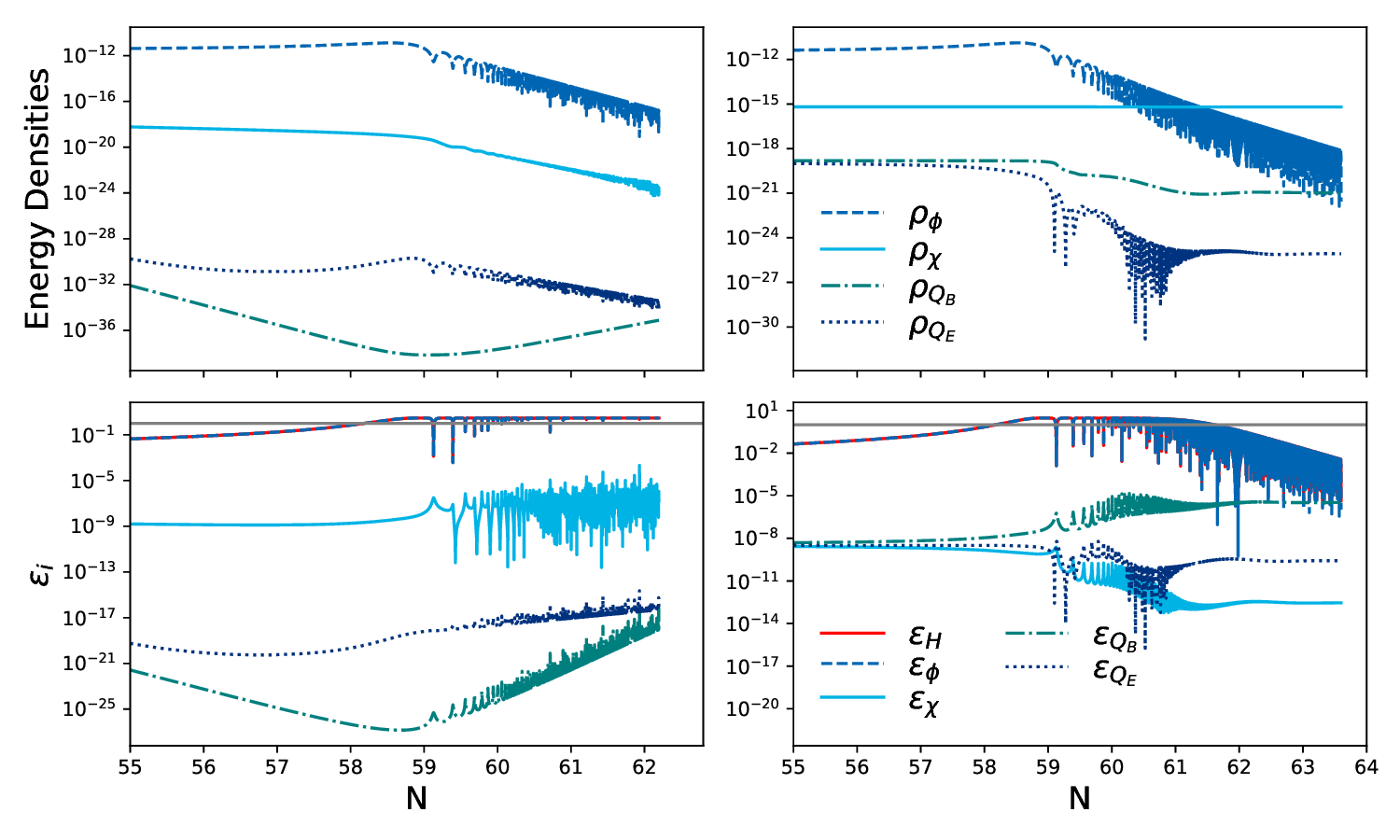}
\caption{The upper panels show the evolution of energy densities of the inflaton field $\rho_{\phi}$ (dashed blue curve), axion field $\rho_{\chi}$ (solid cyan curve), and SU(2) gauge field (dotted dark blue curve for $\rho_{Q_E}$ and dot-dashed green for $\rho_{Q_B}$) for the $\alpha$-attractor potential and runs C1 (left panel) and F (right panel). For the run C1 the axion reaches a minimum
of its potential before the end of inflation. For run F we chose a smaller initial value of the axion field such that a minimum of axion potential is not reached. The latter case leads to the second phase of inflation, dominated by the chromo-natural sector. The lower  panels show the evolution of the first slow-roll parameter $\epsilon_H$, along with the different contributions defined in eq.~\eqref{eq:epsilonH}.} 
\label{energy_densities_alpha_attract}
\end{figure}

Figure \ref{energy_densities_alpha_attract} shows the evolution of the energy densities of the inflaton field $\rho_{\phi}$ (dashed blue curves), axion field $\rho_{\chi}$ (solid cyan curves), and the electric $\rho_{Q_E}$ (dotted dark blue curves) and magnetic $\rho_{Q_B}$ (dot-dashed green curves) components of the background energy density of the SU(2) field, defined in equation \eqref{eq:rho}. Here we use the $\alpha$-attractor potential and runs C1 (left panel) and F (right panel) from table~\ref{table1}. In the lower part of this figure, we show the evolution of the first slow-roll parameter, $\epsilon_H$, along with the different contributions as defined in eq.~\eqref{eq:epsilonH} for these runs.

For run C1, the initial value of the $\chi$ field is $0.975 \pi f$, and the axion reaches a minimum of its potential before the end of inflation, as indicated by the dotted blue curves in  \Fig{Q_and_chi_plot_mu2p69em4_2048a_ct_lhubble}. For run F, we chose a smaller initial value of the $\chi$ field (further away from the minimum and higher up the potential), $\chi/f=0.9 \pi$. For this run, the $\chi$  does not reach a minimum of its potential by the end of inflation and the total energy density of the $\chi$ field remains dominated by its potential energy. Around $N\approx 62$, the total energy budget of the universe becomes dominated by the potential energy of the $\chi$ field, ushering a second inflationary stage, as shown by the evolution of $\epsilon_H$ for run F.
For the remainder of this work, we choose initial parameters for the background axion and gauge field that preclude the existence of a prolonged secondary inflationary stage.

\subsection{Evolution of gauge field modes with vanishing VEV}
\label{sec:endofinflation}
{In this section we examine the evolution of the gauge field modes $T_R$ when the gauge field VEV $Q$ approaches zero. When $m_Q \xi$ becomes zero, \Eq{eq:TRattractor} in terms of conformal time defined as $d\tau=dt/a$ reduces to,
\begin{align}\label{trpostinfl}
    \partial_{\tau}^2 T_R+k^2 T_R=0.
\end{align}
The general solution of the above equation is
\begin{equation}
    T_R=d_1 \sin k\tau+d_2 \cos k \tau.
\end{equation}
By matching this solution to the solution with $\sqrt{2k}\,xT_R=c_1(k)$ at the transition of $Q$ from the backreaction-supported attractor to zero (assuming this is fast enough) in the superhorizon limit ($-k\tau\ll 1$), we obtain the following expression for $T_R$
\begin{align}\label{TLsol}
    T_R\approx\frac{c_1(k)}{\sqrt{2k}}\frac{a_T^2 H^2}{k^2}\Big[\sin k (\tau-\tau_T)+\frac{k}{a_T H}\cos k(\tau-\tau_T)\Big],
\end{align}
where $a_T$ denotes the value of the scale factor at the time when $Q$ transitions  from backreaction-supported attractor to zero during inflation, respectively, and $\tau_T=-1/(a_T H)$.

Let us pause momentarily to discuss this transition. Figure~\ref{tr_and_psi_plot} clearly shows two distinct types of behavior for the gauge field modes $T_R$. For $m_Q\xi\simeq -1$, $T_R\propto 1/x$ and for $m_Q\xi\simeq 0$, $T_R \sim {\rm {const}}$. By using these two simple power-law behaviors, we can define a ``knee'' in the corresponding plot of $T_R$, which for Figure~\ref{tr_and_psi_plot} occurs roughly at $N=45$ $e$-folds. We define the scale-factor at this time as $a_T$. The analysis presented here uses the assumption that this transition is instantaneous. As we see in Figure~\ref{Q_and_chi_plot_mu2p69em4_2048a_ct_lhubble}, the transition of $m_Q\xi$ from $-1$ to $0$ can take a few $e$-folds. However, the introduction of $a_T$ allows us to understand the behavior without unnecessarily complex calculations. Furthermore, in the estimation of the late-time magnetic field that appears in the next section, we use the value of $T_R$ at the end of inflation, as extracted from our full numerical simulation. Therefore we keep the transition scale-factor $a_T$ as a useful notation, keeping in mind the limitations of this approximation.

In the superhorizon limit ($k \tau \ll 1$) the cosine part of eq.~\eqref{TLsol} gives the dominant contribution and the mode function can be approximated as
\begin{align}\label{trconstvalue}
    T_R\approx\frac{c_1(k)}{\sqrt{2k}}\frac{a_T H}{k}.
\end{align}
}
The energy density of $T_R$ after $Q\rightarrow 0$  (equivalently $m_Q\rightarrow 0$) is written in terms of conformal time as
\begin{align}
    \rho_{T_R}=\frac{1}{a^4}\int \frac{d^3 k}{(2\pi)^3}\frac{1}{2}\left(|\partial_{\tau}T_R|^2+k^2|T_R^2|\right).
\end{align}By substituting $T_R$ from \Eq{TLsol} into this expression, we get
\begin{align}\label{ed_of_TR}
  \rho_{T_R}=\frac{1}{4}\left(\frac{a_T}{a}\right)^4 H^4\int \frac{d \ln k}{(2\pi)^3} |c_1(k)|^2 \left(1+\left(\frac{k}{a_T H}\right)^2\right).
\end{align}
From the above expression, we conclude that the energy density of $T_R$ decays as $1/a^4$ after $Q$ becomes zero, as expected for a radiation degree of freedom in an expanding universe.

When $Q\to 0$ before the end of inflation, there is a period during inflation where $T_R$ is almost constant until the transition to $Q=0$ occurs at $a_T$. \Fig{Q_and_chi_plot_mu2p69em4_2048a_ct_lhubble} shows that $Q$ approaches zero between 40 to 50 $e$-folds for run C1 (the dashed blue curve). Consequently, $\sqrt{2k}\, x T_R$ starts decreasing as $x$ decreases, as shown in \Fig{tr_and_psi_plot}. However, if we initially choose a smaller value of $\chi/f$ and a smaller $\mu$ value to
maintain the same value of $m_Q$, the transition of $Q$ to zero happens later compared to run C1. We demonstrate this in \App{appendixa}. Therefore, for fixed $m_Q$ and $g$ values, the largest possible value of $T_R$ at the end of inflation is achieved when the transition of $Q$ to zero happens very close to the end of inflation. In this case, $\sqrt{2k}\,x T_R$ will remain constant until the end of inflation.  In section \ref{sec:magnetogenesis} we  further investigate the  implications for magnetogenesis when $Q$ vanishes at different times during inflation. 

\section{Post inflationary gauge field evolution and magnetic field generation}\label{sec:magnetogenesis}
In this section, we study the evolution of gauge field perturbations after the end of inflation. For the evolution after inflation, we assume that reheating occurs instantaneously, after which the universe transitions into a radiation-dominated era. 
This can be accomplished for example through tachyonic preheating of the inflaton sector. An intriguing possibility is the identification of the inflaton as a pseudo-scalar field (axion) and the natural addition of a $\phi F\tilde F$ coupling of the axion-inflaton to U(1) gauge fields. Since it has been shown \cite{Adshead:2015pva, Adshead:2016iae} that Chern-Simons couplings to U(1) fields can preheat the universe after inflation instantaneously, while leaving the inflationary history largely unaffected (for a proper choice of parameters), this presents a unifying picture of our model, where two axions are coupled to different gauge sectors and one (the inflaton) dominates the energy density and thus drives inflation.

As discussed in the previous section, when $Q$ approaches zero, the solution for $T_R$ is given by \Eq{TLsol}. This solution indicates that $T_R$ remains almost constant when a particular wavelength is much larger than the size of the Hubble horizon and begins oscillating once the mode re-enters the  horizon. In the superhorizon limit, the constant value of $T_R$ is given by ~\Eq{trconstvalue}, which can also be expressed as
\begin{align}
    T_R\approx\frac{c_1(k)}{\sqrt{2k}}\frac{a_T}{a_e}\frac{a_e H}{k}.
\end{align}

Here, $a_e$ denotes the value of the scale factor at the end of inflation, and $a_T/a_e$ accounts for the suppression in the value of $T_R$, depending on how early $Q$ reaches zero before the end of inflation. As $m_Q \xi$ remains zero,  the evolution of $T_R$  after inflation follows eq.~\eqref{trpostinfl}. Therefore, in the post-inflationary era, $T_R$ can be written as an oscillatory function of conformal time
 for modes larger than the Hubble size, while its energy density decays like radiation, as described by \Eq{ed_of_TR}.

Having determined the evolution of the gauge field modes after inflation, we are  ready to consider their evolution through the electroweak phase transition.

\subsection{Magnetogenesis}
\label{subsec:magneto}
At the electroweak era, a component of the SU(2) field transforms into the electromagnetic field. The relation between the electromagnetic field $A_{\mu}$, the SU(2) field $W_{\mu}^a$, and the hypercharge field $B_{\mu}$ is given by
\EQ
A_{\mu}=W_{\mu}^3 \sin{\theta_W}+B_{\mu} \cos{\theta_W},
\EN
where $\theta_W$ is the weak mixing angle (Weinberg angle).
Neglecting the contribution of $B_{\mu}$ and using $\sin{\theta_W}\sim 0.5$, we get
\EQ \label{a&w}
A_{\mu}\approx 0.5 W_{\mu}^3.
\EN

Without loss of generality we associate the perturbations as $\delta W_{3}^{0}=0$,
$\delta W_{i}^{3}=a t_{i3}$. Assuming that tensor perturbations of the $SU(2)$ gauge field give the dominating contribution after inflation, from eqs.~\eqref{eq:decomposition}, \eqref{eq:tijhij} and \eqref{eq:thnormalised} we arrive at
\begin{align}
|A_{i}|^2&=\frac{1}{4}\left(|T_{L}|^2+|T_{R}|^2\right).\label{A_in_TL}
\end{align}
\begin{figure}[h!]
\centering
\includegraphics[width=1\textwidth]{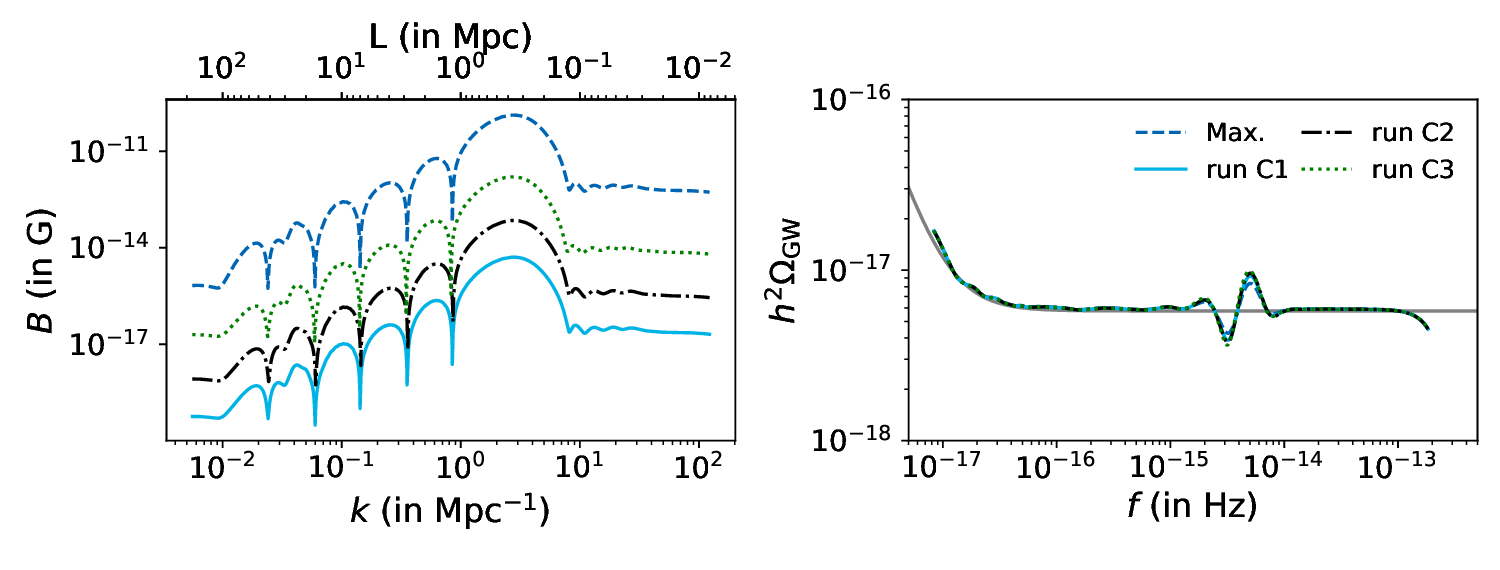}
\caption{The left panel of this figure shows the magnetic field amplitude $B$ at the present epoch given by \Eq{mf_estimate} with respect to the wave number and coherent length. We use the $\alpha$-attractor potential and runs C1 (solid cyan curve), C2 (dot-dashed black) and C3 (dotted green) from table~\ref{table1}. The dashed blue curve corresponds to run C1 with $a_T/a_e=1$, which implies that $\sqrt{2k}\, x T_R$ for all the super-Hubble modes remains constant until the end of inflation. The evolution of $Q$ and $\chi/f$ for these runs is shown in \Fig{Q_and_chi_plot_g0p1} in appendix~\ref{appendixa} and
as a dashed blue curve in \Fig{Q_and_chi_plot_mu2p69em4_2048a_ct_lhubble} for run C1. The right panel shows the corresponding gravitational wave spectral energy density fraction with the same color coding and its vacuum contribution (estimated using eq.~\eqref{gw_vacuum}) in the grey curve.}
 \label{omegagw_plot}
\end{figure}
In terms of the vector potential, $A_i$, the magnetic energy spectrum is given by (see eq.~(17) in ref.~\cite{Sharma:2018kgs}),
\begin{align}
\Delta_{\rm B}(k)&=\frac{1}{(2\pi)^2} \frac{k^5}{a^4} |A_i|^2.
\end{align}
Here, $\Delta_{\rm B}$ represents the magnetic energy spectrum per logarithmic wave number interval 
and is defined such that the magnetic energy density is $\rho_B\equiv\langle B^2/2\rangle=\int d \ln k \,\Delta_{\rm B}(k)$.
Using \Eqs{TLsol}{A_in_TL}, we arrive at the following expression for the magnetic energy spectrum at the EW epoch
\begin{align}
\Delta_{\rm B}(k)&=\frac{1}{(2\pi)^2} \frac{k^5}{a^4}\frac{1}{4}\left(|T_{L}|^2+|T_{R}|^2\right)\sim\frac{1}{(2\pi)^2} 
\frac{a_e^4 H_e^4}{a^4} \left(\frac{a_T}{a_e}\right)^4\frac{1}{8} |c_1|^2 \sin^2 k (\tau-\tau_e).
\end{align}
Here $H_e$ represents the value of the Hubble parameter at the end of inflation. For  subhorizon modes ($k\tau\gg1$), the typical value of $\sin^2 k (\tau-\tau_e)$ can be approximated by 1/2.
Furthermore, by normalizing $\Delta_{\rm B}(k)$ with the total energy density of the universe at the end of inflation, $\rho_e=3 H_e^2 M_{\rm pl}^2$, we can write
\begin{align}
\frac{\Delta_{\rm B}(k)}{\rho_e}&\sim\frac{a^4}{3 H_e^2 M_{\rm pl}^2 a_e^4}\frac{1}{16(2\pi)^2} 
\frac{a_e^4 H_e^4}{a^4}\left(\frac{a_T}{a_e}\right)^4|c_1|^2=\frac{1}{48(2\pi)^2} 
\left(\frac{H_e}{M_{\rm pl}}\right)^2\left(\frac{a_T}{a_e}\right)^4|c_1|^2.
\end{align}
The above expression implies that the magnetic energy spectrum is proportional to $|c_1|^2$. Using the value of the radiation energy density at the present epoch to be $\sim (3\,\mu{\rm G})^2$, the magnetic field strength at its peak wave number becomes
\begin{align}
B\approx \sqrt{2\Delta_{\rm B}(k)\big|_0}&=
9.7 \times 10^{-8}
\frac{H_e}{10^{-6} M_{\rm pl}}\left(\frac{a_T}{a_e}\right)^2|c_1|~ \mu{\rm G}\label{mf_estimate}.
\end{align}

We use equation \eqref{mf_estimate} to compute the magnetic field strength at the present epoch, using the value of $(a_T/a_e)^2 |c_1|$ obtained from the simulation at the end of inflation, as discussed earlier. The resulting amplitude with respect to the wave number $k$ and the corresponding length scale are shown in the left panel of \Fig{omegagw_plot} for runs C1--C3. The peak value of the obtained magnetic field strength is $5.3 \times 10^{-15}~\rm G$, $7.4 \times 10^{-14}~\rm G$, and $1.6 \times 10^{-12}~\rm G$ for the runs C1 (solid cyan curve), C2 (dot-dashed black curve), and C3 (dotted green curve), respectively. The dashed blue curve represents the case where we used  $a_T=a_e$  in \Eq{mf_estimate} for run C1 and the corresponding magnetic field strength is $1.3 \times 10^{-10}~\rm G$. This occurs when the initial value of $\chi$ is fine-tuned, so that $Q\to 0$ very close to the end of inflation. As can be inferred from  table~\ref{table1}, such fine-tuning requires choosing the initial value of $\chi/f$ at the $0.1\%$ level. Since this is not necessary for the viability of our model, we do not attempt to provide this exact value.
Therefore, to obtain the magnetic field strength shown in the dashed blue curve, we use the value of $|c_1|$ from run C1 at $N=30$, where $x |T_R|$ is in the regime where it is almost constant in time.

\begin{figure}
\centering
\includegraphics[width=0.55\textwidth]{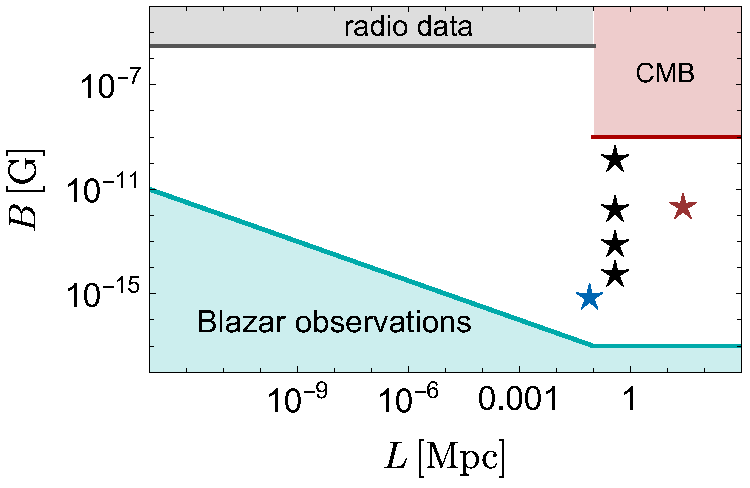}
\caption{Bounds on intergalactic magnetic fields adapted from reference \cite{MAGIC:2022piy}. The light blue-shaded region shows the lower bound inferred from blazar observations \cite{Taylor:2011bn}, the red-shaded upper bound shows Planck Collaboration analyses \cite{Planck:2015zrl} and
the light-grey shaded upper bound are conservative limits from radio data \cite{Kronberg:1993vk}  and theoretical estimates \cite{Durrer:2013pga}. By black stars we illustrate the magnetic field amplitudes from figure \ref{omegagw_plot} and the red and blue stars
 refer to corresponding peaks of the red-dashed and blue curves in figure \ref{fig:GWs}. }
 \label{bounds}
\end{figure}

The magnetic energy spectra peak at a length scale of approximately $0.4~\rm Mpc$ for these cases
with amplitudes that satisfy the lower bound from blazar observations, as shown in figure~\ref{bounds} with black stars. The red star on the figure represents the case with the larger value of $m_Q$. In that situation the backreaction effects become important earlier and the system transitions to the backreaction-supported attractor shortly after the start of inflation, moving the peak of magnetic energy spectra to larger scales, as discussed in appendix \ref{appendix:GW} (see the corresponding red-dashed curve in figure \ref{fig:GWs}). For smaller $m_Q$ values, this transition happens later, pushing the magnetic field peak value to smaller scales, as represented by the blue star and corresponding to the blue curve in figure \ref{fig:GWs}. The resulting amplitude of the magnetic field depends on the initial value of parameter $m_Q$ as well on how close to the end of inflation the gauge field VEV converges to zero. From figure~\ref{bounds} it follows that the magnetic fields produced during spectator chromo-natural inflation can potentially {explain} the presence of the magnetic fields in the intergalactic medium. 

In \Fig{omegagw_plot}, the wave number $k$ at the present epoch is computed as
\begin{equation}
k=e^{-(N_e-N_k)}k_H,
\end{equation}
where $N_e$ and $N_k=\ln (k/H)$ represent the total number of $e$-folds during inflation and the number of $e$-folds at which the wave number $k$ exits the Hubble horizon during inflation, respectively, and $k_H$ represents the present-day value of the wave number corresponding to the Hubble size at the end of inflation and is given by
\begin{align}
k_H=\frac{a_e}{a_0} H=2.3 \times 10^{22} ~\rm Mpc^{-1} \left(\frac{H}{10^{-6}M_{\rm pl}}\right)^{1/2}.
\end{align}
In the above expression, we assumed an adiabatic evolution of the universe,
\begin{align}
\frac{a_e}{a_0} = \left(\frac{g_{0s}}{g_{rs}}\right)^{1/3}\frac{T_0}{T_r} = 5.96 \times 10^{-29} \left(\frac{g_{0s}}{3.94} \frac{106.75}{g_{rs}}\right)^{1/3} \frac{T_0}{2.73 \text{K}} \left(\frac{ 10^{-6} M_{\rm pl}}{H}\right)^{1/2},
\end{align}
where $g_{rs}$ and $g_{0s}$ denote the effective
degrees of freedom in the entropy at the end of inflation and the present
epoch, respectively. We estimate the reheating temperature, $T_r$, by assuming instantaneous reheating using
$3H^2M_\text{pl}^2 = (\pi^2/30)g_rT_r^4$.

Furthermore, we calculate the gravitational wave spectral energy density fraction, $\Omega_{\rm GW}h^2$  defined as 
\begin{equation}\label{omegagw}
h^2\Omega_{\rm GW}(k)=\frac{3}{128}h^2\Omega_{\rm rad}{\cal P}_h^{\rm tot}(k)\left[ \frac{1}{2}\left(\frac{k_{\rm eq}}{k}\right)^2+\frac{16}{9}\right].
\end{equation}
Here, $h^2\Omega_{\rm rad}\simeq 2.47 \times 10^{-5}$ represents the current radiation density fraction, and $k_{\rm eq} \simeq 1.3 \times 10^{-2}~\rm Mpc^{-1}$ is the wave number corresponding to the Hubble horizon at matter-radiation equality. The parameter $h$ is defined such that
$H_0 = 100\, h \, {\rm km} \, {\rm s}^{-1} ~\rm Mpc^{-1}$,
where $H_0$ is the Hubble parameter at the present epoch. To express $\Omega_{\rm GW}$ in terms of frequency, $f$ instead of wave number, $k$, we use $f \simeq 1.5 \times 10^{-15} (k/1~\rm Mpc^{-1}) \, {\rm Hz}$. In the expression \eqref{omegagw} ${\cal P}_h^{\rm tot}(k)$ is the total power spectrum of sourced gravitational waves by the tensor perturbations of the SU(2)-gauge field, defined as
\begin{equation}
{\cal P}_h^{\rm tot}(k)=\frac{{2}H^2}{\pi^2 M_\text{pl}^2}\sum_{s=L,R}\Bigl|\sqrt{2k}\,\left(\frac{k}{a H}\right) \lim_{k/aH\rightarrow 0} \psi_{(s)}\Bigr|^2
\end{equation}
for the sourced contribution from the tensor perturbations of the SU(2)-gauge field.
The vacuum contribution of tensor the metric perturbations is 
\begin{equation}\label{gw_vacuum}
{\cal P}_h^{\rm vac}(k)=\frac{2 H^2}{\pi^2 M_\text{pl}^2}.
\end{equation}
We show the 
gravitational wave spectral energy density fraction and its comparison to vacuum contribution in the right panel of \Fig{omegagw_plot}.  We can see that oscillations in $T_R$ produce oscillations in $\Omega_{\rm GW}h^2$, but for the runs C1--C3, the amplitude of gravitational waves is small and unobservable in the upcoming surveys. For larger values of $m_Q$, the amplification is significant. We show this in appendix~\ref{appendix:GW}.

\subsection{Comparison with  magnetogenesis from axion-U(1) inflation}

It is important to compare  the underlying physics of magnetogenesis from axion-U(1)  inflation~\cite{Carroll:1991zs, Fujita:2015iga,Adshead:2016iae,Gorbar:2021zlr} to our current work. In the case of axion-U(1) inflation, one of the gauge field modes is amplified due to the coupling between the axion and U(1). The strength of this amplification depends on the velocity of the axion, and the axion-gauge coupling—a faster-rolling axion results in more rapid growth of the gauge field. As the axion’s velocity becomes maximal near and after  the end of inflation, the gauge field modes with wavelength comparable to the horizon at this time experience maximum amplification. In practice, the most efficient amplification of gauge fields occurs during preheating, where it was shown in ref.~\cite{Adshead:2016iae} that the inflaton can transfer the entirety of its energy density to gauge fields, leading to a magnetic field strength $B^2\sim M_{\rm Pl}^2 H^2$. 
Consequently, a spectrum is obtained that peaks around the Hubble horizon scale near the end of inflation with large amplitude.

These fields are largely helical\footnote{The gauge fields would be exponentially close to being totally helical, but non-rescattering during preheating alleviates part of the helicity~\cite{Adshead:2016iae}.} and undergo an inverse cascade, leading to typical length scales on the order of parsecs, with a strength of around $10^{-13}~\rm G$ \cite{Adshead:2016iae}. The wave number modes corresponding to present-day length scales of approximately $1~\rm Mpc$ leave the horizon about 10 $e$-folds into inflation. However, their amplitude continues to decay even after crossing the horizon, resulting in a very small magnetic field at the end of inflation, leading to tiny magnetic field strength at those scales.

In contrast, the dynamics in the case of inflation with a (spectator) axion-SU(2) sector is quite different due to the existence of the backreaction-supported attractor. Since the magnetic field arises from a component of the SU(2) gauge field’s tensor perturbations, the magnetic field spectrum is related to the tensor perturbation spectrum. The tensor perturbation spectrum peaks roughly at a scale corresponding to the Hubble horizon scale around the epoch when $Q$ transitions from the initial spectator chromo-natural attractor to the backreaction-supported attractor. As an example, in the case of runs C1--C3, this transition occurs around $10$ $e$-folds from the start of inflation; see figure \ref{Q_and_chi_plot_mu2p69em4_2048a_ct_lhubble}. Therefore the  modes which exit the Hubble horizon around this time experience maximum amplification. The backreaction from these perturbations becomes important and the system transitions to the backreaction-supported attractor. During this stage,  $m_Q \xi \simeq -1$, which leads to a constant amplitude of  $\sqrt{2k}\,x T_R$ during the super-horizon evolution, with roughly no decay until $Q$ reaches zero. Modes that exit the horizon during the backreaction-supported attractor phase do not undergo much amplification due to the small value of $m_Q$. Referring again to runs~C1--C3, backreaction becomes significant around $10$ $e$-folds, and the modes that exit around this time, or earlier, correspond to a length scale of the order of $1~\rm Mpc$ at the present epoch. This is why the magnetic field spectrum shown in the \Fig{omegagw_plot} peaks at megaparsec scales. Simply put, the Abelian case relies on extremely efficient energy transfer to gauge fields, albeit at small scales, whereas the non-Abelian case relies on the non-decay of gauge fields during the back-reaction supported attractor, allowing for much larger correlation lengths, albeit with weaker field strength.

It is instructive to compare the dynamics of the backreaction-supported attractor with the Anber-Sorbo solution \cite{Anber:2009ua}, which describes the homogeneous backreaction regime in axion-U(1) inflation and has recently been shown to be unstable \cite{vonEckardstein:2023gwk}. While in both U(1) and SU(2) cases the backreaction is treated similarly through a homogeneous Hartree-type approximation, the dynamics differ significantly due to the presence of a non-zero VEV in the non-Abelian case. 
The Anber-Sorbo solution for the U(1) case hinges on the continuous production of gauge field modes with ever increasing comoving wave number and thus a change in their growth rate can destabilize it.
Conversely, in the SU(2) case, the primary contribution to the backreaction integrals arises only from a fixed range of comoving wave numbers, which allows for a stable solution.
We refer the interested reader to Appendix D of ref.~\cite{Iarygina:2023mtj} for a detailed discussion on this comparison. Though the backreaction-supported attractor in the non-Abelian case breaks down close to the end of inflation, where the solution smoothly converges to zero, it provides an excellent approximation during inflation.

\subsection{Magnetic mass effects}

Non-Abelian gauge bosons  in a high-temperature plasma, as the one present in the early universe, 
 can acquire an additional mass, dubbed  ``magnetic mass''.  In the previous discussion, we did not consider effects coming from a magnetic mass, which is typically considered to be of the form $m_{\rm mag}\propto$ $ g^2 T$ \cite{Gross:1980br}, where $T$ is the temperature of the universe and $g$ is the gauge-field coupling. The effect of the additional mass term can be estimated using \cite{Maleknejad:2020pec}
\begin{equation}
     \partial_{\tau}^2 T_R+\left(k^2 +a^2 m^2_{\rm mag}\right)T_R\simeq 0.
\end{equation}
However, a field with a mass proportional to the temperature behaves like radiation, meaning that the magnetic field scaling shown in section~\ref{sec:magnetogenesis} is still valid. Since in the radiation-dominated era $a \propto \tau$ and $T\propto \sqrt{H}\propto \frac{1}{\tau}$, leading to $a^2 m^2_{\rm mag} = {\rm {const}}$, the magnetic mass will modify the oscillation frequency of perturbations, leaving the amplitude unchanged (except possible  effects from the time-varying initial creation of the thermal plasma).

An interesting analogy of the EWPT is that of a superconductor.
We can imagine the super-horizon gauge fields before the EWPT as a magnetic field permeating a superconductor with a temperature larger than the superconducting phase transition. If one lower the temperature, when the solid turns into a superconductor, the previously homogeneous magnetic field will break into filaments surrounded by current vortices. 
Similar formation of magnetic structures,  including flux tubes, filaments and vortices, also occurs when magnetic fields interact with a plasma. It is intriguing to further explore the details of the magnetic field evolution through the EWPT. While it is computationally challenging, progress in simulating structures in the full electroweak theory has been made (see e.g.~\cite{Patel:2023ybi}) and such a simulation is necessary to probe 
 whether magnetic fields produced in this model will acquire   spatial patterns when the Higgs relaxes to its VEV and the SU(2) fields get partially transformed into electromagnetic fields.

\section{Summary and Discussion}\label{sec:conclusions}

We have explored the consequences of spectator chromo natural inflation (SCNI), where the non-Abelian gauge field is identified with the SU(2)$_{\rm L}$ sector of the Standard Model. This fixes the gauge coupling to be large $g\gtrsim {\cal O}(0.1)$, bringing the backreaction contribution from tensor perturbations to be comparable to other terms in the background equations of motion. The system necessarily flows into the recently discovered backreaction-supported attractor that appears in this regime \cite{Iarygina:2023mtj}.  

{In our previous work \cite{Iarygina:2023mtj}, we studied the evolution of the SCNI sector during inflation 
under the assumption of a constant Hubble parameter and discovered a new type of dynamical attractor, supported by the back-reaction of gauge field fluctuations on the background trajectory. Here, we relaxed this assumption and analyzed the dynamics of the axion-gauge field system until the end of inflation, using quadratic and alpha-attractor potentials to model the  background evolution of the inflaton.
We stressed an important point, which was overlooked so far in the literature. In these models there is a possibility for a second inflationary stage, after the inflaton rolls to its potential minimum, driven by the energy density of the axion field of the SCNI sector. To avoid this case, we focused on cases where the initial value of the axion field is chosen such that it reaches the minimum of its potential before the end of inflation. When the axion field approaches a minimum of its potential, the  gauge field VEV smoothly transitions to zero, and the tensor perturbations of the gauge fields begin to red-shift as expected for gauge fields in an FRW spacetime. During the EWPT, part of the tensor perturbations {{of the SU(2) gauge field}} get transformed into the electromagnetic part of the broken SU(2)$_{\rm L}\times $ U(1)$_{\rm Y}$ sector.
The electric component of the fields will be quickly damped (typically within one Hubble time) due to the large conductivity of the Universe. However, the magnetic component will remain frozen,  providing a viable origin for the presence of magnetic fields in the intergalactic medium. The obtained magnetic field at the present epoch depends on the axion-SU(2) model parameters. For one set of parameter choices presented here, we found that the magnetic fields have a strength of $5 \times 10^{-15}~\rm G$ with
a coherence length of approximately $0.4~\rm Mpc$ at the present epoch. This is above the lower bound on the strength of the magnetic field in the intergalactic medium inferred from GeV observations of blazars.

Given the intriguing dynamics and important phenomenology of this model, several avenues for future work arise.
So far, we have analyzed the dynamics of the axion-SU(2) system using the linear evolution equations for the gauge field modes. It was recently shown~\cite{Dimastrogiovanni:2024lzj} that accounting for gauge field self-interactions and axion-gauge
field non-linear couplings leads to  bounds on the parameters of the model, so that a perturbative description of the theory  is valid. Interestingly, these bounds on the parameter space of the theory are comparable to the edge of the strong backreaction regime. It is thus necessary to perform a full numerical computation to accurately determine the exact perturbativity bounds and their competition with the strong backreaction regime. Moreover, our analysis neglects spatially dependent backreaction
effects, that have been shown to have a strong impact on the overall dynamics close to the end of inflation in the Abelian case~\cite{Figueroa:2023oxc}. Performing lattice simulations would be a natural next step to explore the non-linearities in axion-SU(2) gauge field dynamics.
Finally, the detailed evolution of the produced SU(2) gauge fields through the EWPT and the possibility of the creation of magnetic field filaments, akin to the case of a superconductor, is beyond the scope of our present calculation.
Further analysis of these exciting aspects is left for future work.

\section*{Acknowledgments}
We thank G.~Dvali, T.~Fujita, K.~Kamada, A.~Long, K.~Mukaida, K.~Subramanian and T.~Vachaspati for useful discussions on the evolution of non-Abelian fields. 
A.B., O.I. and E.I.S.\ acknowledge the hospitality of the Bernoulli Center  during the workshop ``Generation, evolution, and observations of cosmological magnetic fields", 
where part of this work was conducted and presented. A.B.\ was supported in part by the Swedish Research Council (Vetenskapsr{\aa}det) under Grant No.\ 2019-04234,
the National Science Foundation under Grants No.\ NSF PHY-2309135 and AST-2307698, and the NASA ATP Award 80NSSC22K0825.
The work of O.I.\ was supported by the European Union's Horizon 2020
research and innovation program under the Marie Skłodowska-Curie grant
agreement No.~101106874. R.S.\ was supported by the Czech Science Foundation (GAČR), project 24-13079S.
We acknowledge the allocation of computing resources provided by the Swedish National Allocations Committee at the Center for Parallel Computers at the Royal Institute of Technology in Stockholm.

\paragraph{Data availability.}
The source code used for the numerical solutions of this study, the
{\sc Pencil Code}, along with the module \texttt{special/axionSU2back} used in the present study, are freely available at \url{https://github.com/pencil-code/pencil-code/}.
The numerical data and input files are available on \url{http://norlx65.nordita.org/~brandenb/projects/magnetogenesis-SU2/}.

\appendix

\section{Different initial values of $\chi/f$ and $\mu$}\label{appendixa}
In this section we show the evolution of the axion-SU(2) system for different initial values of $\chi/f$ and $\mu$, while keeping the initial values of $m_Q$ constant for a fixed value of $g$. In \Fig{Q_and_chi_plot_g0p1} we illustrate the evolution of $Q$ and $\chi/f$ for runs C1 (solid blue curves), C2 (dashed black curves) and C3 (dotted green curves). The initial parameters for these runs are provided in \Tab{table1}. 
\begin{figure}
\centering
\includegraphics[width=1\textwidth]{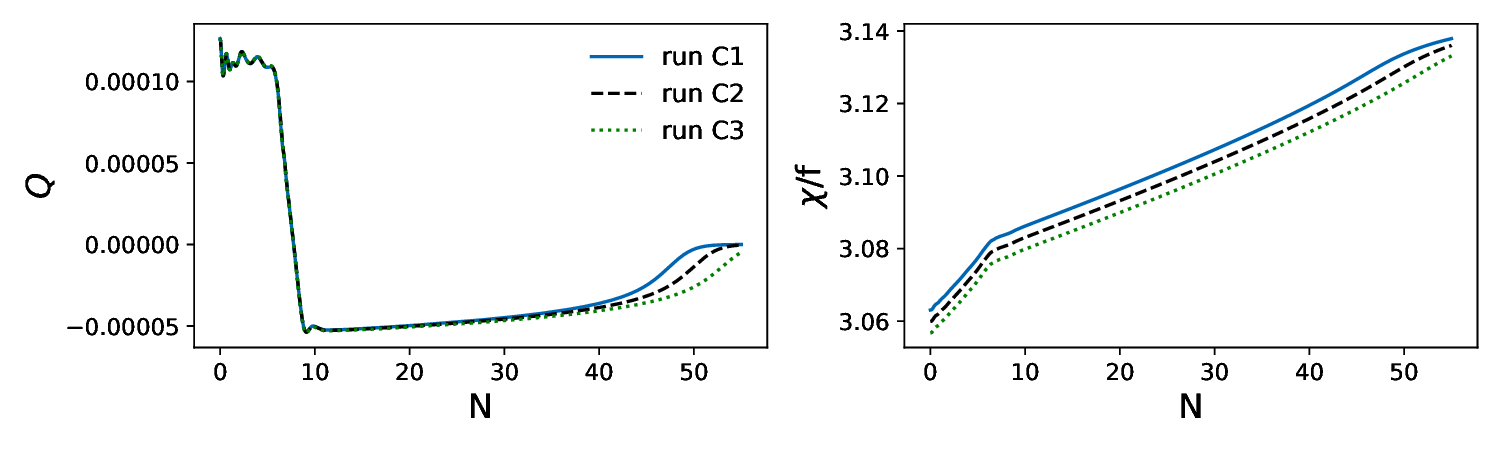}
\caption{The evolution of the $Q$ (left panel) and $\chi/f$ (right panel) for the runs C1 (solid blue curves), C2 (dashed black curves), and C3 (dotted green curves).}
\label{Q_and_chi_plot_g0p1}
\end{figure}

From \Fig{Q_and_chi_plot_g0p1}, we conclude that by choosing a smaller initial value of $\chi/f$ with the same initial value of $m_Q$, the transition of $Q$ to zero occurs later. This behavior impacts the value of tensor perturbations $T_R$ at the end of inflation, leading to a larger $T_R$ if the transition happens later. Consequently, the resulting  magnetic field value at the present epoch will also be larger, as demonstrated in \Fig{omegagw_plot}.

\section{Gravitational waves and magnetic fields for higher gauge field couplings}\label{appendix:GW}
The amplitude of GWs, we computed in section \ref{subsec:magneto} is approximately equal to their vacuum contribution. This happens because gauge field amplification is not sufficient to source metric tensor perturbations. However, for larger values of $m_Q$, the amplification becomes higher, leading to a sourced contribution of GWs that exceeds the vacuum value. In figure~\ref{fig:GWs} we demonstrate the amplification of gravitational waves for higher values of $m_Q$. In this figure,  we compare the results of run C1 from figure \ref{omegagw_plot} (black dotted curves, with $m_Q= 2.37$)  to run G (dashed red curves, with $m_Q= 5.71$) and run H (solid blue curves, with $m_Q=2.34$) from table \ref{table1}. The top left panel shows the magnetic field strength and the top right the corresponding amplification of gravitational waves. The background evolution of $Q$ and $\chi/f$ is illustrated in the bottom left and right panels, respectively. For higher values of $m_Q$ the backreaction effects become important earlier, forcing the system into the new backreaction-supported attractor very close to the start of our simulation. This leads to the amplification of tensor perturbations on larger scales, with the resulting peak of magnetic energy spectra at a length scale of order $10~\rm Mpc$. This shifts the peak of GWs signal towards smaller frequencies. We avoid taking higher values of $m_Q$, keeping in mind the constraints from bounds on perturbativity. We note that this scenario with $m_Q=5.71$ is already well within the backreaction regime and may conflict with perturbativity bounds. Here, we aim to demonstrate how the amplification of GWs might still be achieved. However, a thorough study of such scenarios would require lattice simulations of the axion-SU(2) system.

\begin{figure}[h!]
\centering
\includegraphics[width=1\textwidth]{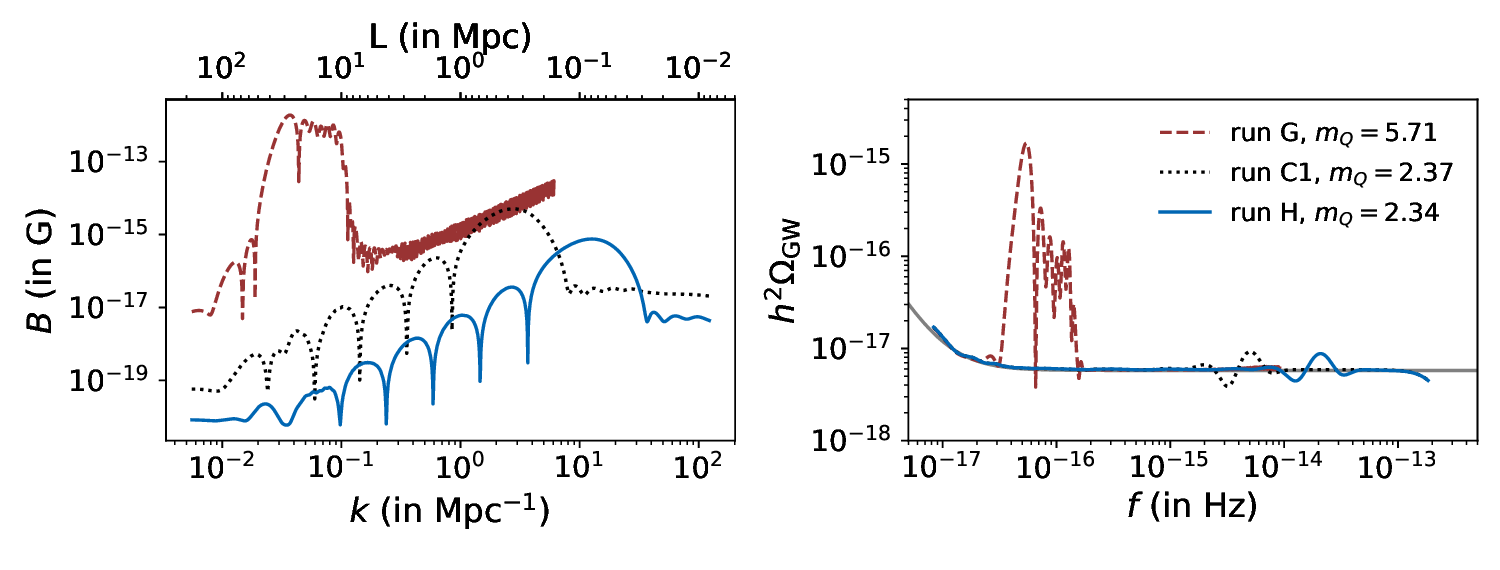}
\includegraphics[width=1\textwidth]{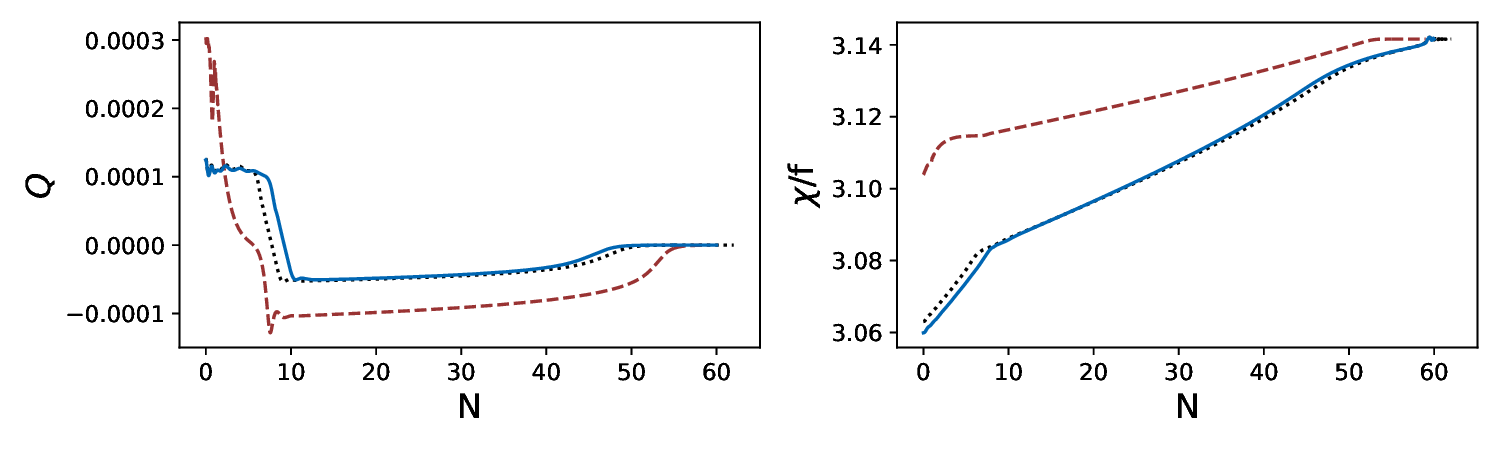}
\caption{In this figure, we show the obtained magnetic field strength at the present epoch and the GW density fraction $\Omega_{\rm GW}h^2$ in the top panels for the run G (dashed red curves), run C1 (black dotted curves) and run H (solid blue curves). The value of $m_Q$ is 5.71 for run G, 2.37 for run C1 and for 2.34 run H. Bottom panels show the background evolution of $Q$ and $\chi/f$.}
 \label{fig:GWs}
\end{figure}

It is worth noting that larger values of $m_Q$ can be also achieved by choosing larger gauge field couplings. For bigger couplings,
the dynamics is similar to the case with $g=0.1$ considered in the paper, meaning that our computation is still valid for the case $g=0.65$. The difference is that for $g=0.65$ the backreaction of tensor perturbations on the background evolution is significant right away from the beginning of inflation, even for the smallest allowed value of $m_Q$. 
Figure \ref{Q_and_chi_plot_g0p65} shows two runs with $g=0.65$ and different initial values of $\chi/f=0.975 \pi$ (run D) and $0.973 \pi$ (run E), with the value of $\mu$ chosen such that the initial value of $m_Q$ stays the same. As demonstrated in figure \ref{Q_and_chi_plot_g0p65}, the dynamics is similar to the evolution in figures \ref{Q_and_chi_plot_mu2p69em4_2048a_ct_lhubble} and \ref{fig:GWs}, but $Q$ transits to the backreaction-supported attractor solution at the very early stages of inflation.

\begin{figure}[h!]
\centering
\includegraphics[width=1\textwidth]{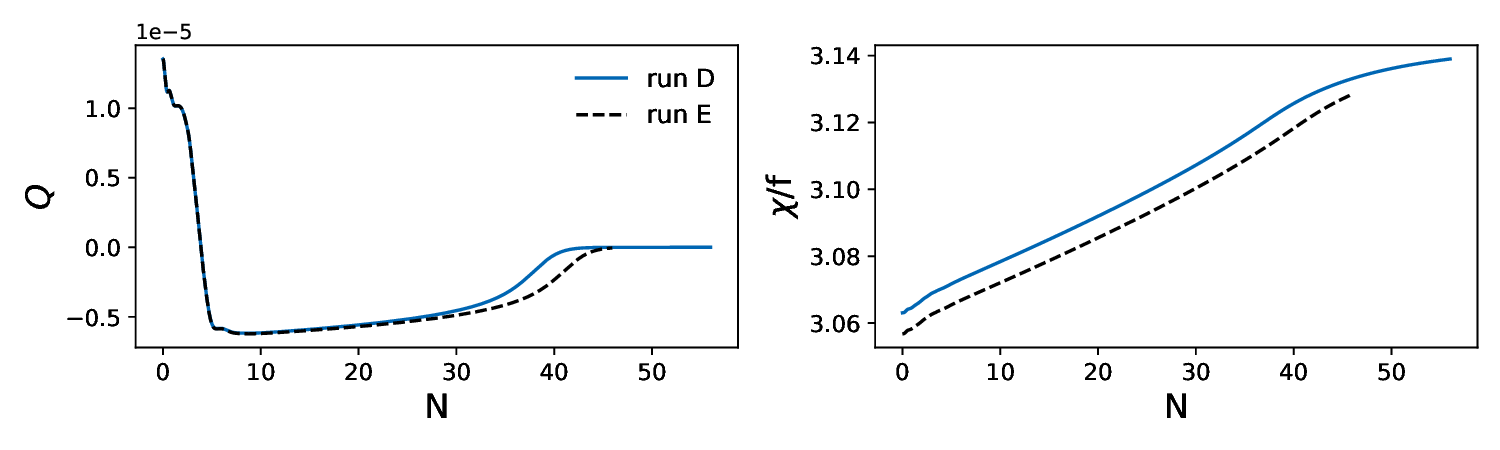}
\caption{In this figure, we show the evolution of the $Q$ and $\chi/f$ for the run D (solid blue curves), and E (dotted black curves). These runs are for the case $g=0.65$.}
\label{Q_and_chi_plot_g0p65}
\end{figure}

\bibliographystyle{bibi}
\bibliography{refs}

\providecommand{\href}[2]{#2}\begingroup\raggedright\begin{thebibliography}{100}

\bibitem{Widrow:2002ud}
L.~M. Widrow, \emph{{Origin of galactic and extragalactic magnetic fields}},
  \href{https://doi.org/10.1103/RevModPhys.74.775}{\emph{Rev. Mod. Phys.}
  {\bfseries 74} (2002) 775}
  [\href{https://arxiv.org/abs/astro-ph/0207240}{{\ttfamily
  astro-ph/0207240}}].

\bibitem{Bernet:2008qp}
M.~L. Bernet, F.~Miniati, S.~J. Lilly, P.~P. Kronberg and
  M.~Dessauges-Zavadsky, \emph{{Strong magnetic fields in normal galaxies at
  high redshifts}}, \href{https://doi.org/10.1038/nature07105}{\emph{Nature}
  {\bfseries 454} (2008) 302}
  [\href{https://arxiv.org/abs/0807.3347}{{\ttfamily 0807.3347}}].

\bibitem{Beck:2008ty}
R.~Beck, \emph{{Galactic and Extragalactic Magnetic Fields}},
  \href{https://doi.org/10.1063/1.3076806}{\emph{AIP Conf. Proc.} {\bfseries
  1085} (2009) 83} [\href{https://arxiv.org/abs/0810.2923}{{\ttfamily
  0810.2923}}].

\bibitem{Kronberg:2007dy}
P.~P. Kronberg, M.~L. Bernet, F.~Miniati, S.~J. Lilly, M.~B. Short and D.~M.
  Higdon, \emph{{A Global Probe of Cosmic Magnetic Fields to High Redshifts}},
  \href{https://doi.org/10.1086/527281}{\emph{Astrophys. J.} {\bfseries 676}
  (2008) 7079} [\href{https://arxiv.org/abs/0712.0435}{{\ttfamily 0712.0435}}].

\bibitem{Clarke_2001}
T.~E. Clarke, P.~P. Kronberg and H.~Böhringer, \emph{A new radio-x-ray probe
  of galaxy cluster magnetic fields},
  \href{https://doi.org/10.1086/318896}{\emph{The Astrophysical Journal}
  {\bfseries 547} (2001) L111}.

\bibitem{Govoni:2004as}
F.~Govoni and L.~Feretti, \emph{{Magnetic field in clusters of galaxies}},
  \href{https://doi.org/10.1142/S0218271804005080}{\emph{Int. J. Mod. Phys. D}
  {\bfseries 13} (2004) 1549}
  [\href{https://arxiv.org/abs/astro-ph/0410182}{{\ttfamily
  astro-ph/0410182}}].

\bibitem{2005A&A...434...67V}
C.~{Vogt} and T.~A. {En{\ss}lin}, \emph{{A Bayesian view on Faraday rotation
  maps Seeing the magnetic power spectra in galaxy clusters}},
  \href{https://doi.org/10.1051/0004-6361:20041839}{\emph{Astronomy and
  Astrophysics} {\bfseries 434} (2005) 67}
  [\href{https://arxiv.org/abs/astro-ph/0501211}{{\ttfamily
  astro-ph/0501211}}].

\bibitem{Neronov:1900zz}
A.~Neronov and I.~Vovk, \emph{{Evidence for strong extragalactic magnetic
  fields from Fermi observations of TeV blazars}},
  \href{https://doi.org/10.1126/science.1184192}{\emph{Science} {\bfseries 328}
  (2010) 73} [\href{https://arxiv.org/abs/1006.3504}{{\ttfamily 1006.3504}}].

\bibitem{Tavecchio:2010mk}
F.~Tavecchio, G.~Ghisellini, L.~Foschini, G.~Bonnoli, G.~Ghirlanda and
  P.~Coppi, \emph{{The intergalactic magnetic field constrained by Fermi/LAT
  observations of the TeV blazar 1ES 0229+200}},
  \href{https://doi.org/10.1111/j.1745-3933.2010.00884.x}{\emph{Mon. Not. Roy.
  Astron. Soc.} {\bfseries 406} (2010) L70}
  [\href{https://arxiv.org/abs/1004.1329}{{\ttfamily 1004.1329}}].

\bibitem{Dolag:2010ni}
K.~Dolag, M.~Kachelriess, S.~Ostapchenko and R.~Tomas, \emph{{Lower limit on
  the strength and filling factor of extragalactic magnetic fields}},
  \href{https://doi.org/10.1088/2041-8205/727/1/L4}{\emph{Astrophys. J. Lett.}
  {\bfseries 727} (2011) L4} [\href{https://arxiv.org/abs/1009.1782}{{\ttfamily
  1009.1782}}].

\bibitem{Essey:2010nd}
W.~Essey, S.~Ando and A.~Kusenko, \emph{{Determination of intergalactic
  magnetic fields from gamma ray data}},
  \href{https://doi.org/10.1016/j.astropartphys.2011.06.010}{\emph{Astropart.
  Phys.} {\bfseries 35} (2011) 135}
  [\href{https://arxiv.org/abs/1012.5313}{{\ttfamily 1012.5313}}].

\bibitem{Taylor:2011bn}
A.~M. Taylor, I.~Vovk and A.~Neronov, \emph{{Extragalactic magnetic fields
  constraints from simultaneous GeV-TeV observations of blazars}},
  \href{https://doi.org/10.1051/0004-6361/201116441}{\emph{Astron. Astrophys.}
  {\bfseries 529} (2011) A144}
  [\href{https://arxiv.org/abs/1101.0932}{{\ttfamily 1101.0932}}].

\bibitem{Takahashi:2013lba}
K.~Takahashi, M.~Mori, K.~Ichiki, S.~Inoue and H.~Takami, \emph{{Lower Bounds
  on Magnetic Fields in Intergalactic Voids from Long-term GeV-TeV Light Curves
  of the Blazar Mrk 421}},
  \href{https://doi.org/10.1088/2041-8205/771/2/L42}{\emph{Astrophys. J. Lett.}
  {\bfseries 771} (2013) L42}
  [\href{https://arxiv.org/abs/1303.3069}{{\ttfamily 1303.3069}}].

\bibitem{Finke:2013tyq}
{\scshape Fermi-LAT} Collaboration, J.~Finke, L.~Reyes and M.~Georganopoulos,
  \emph{{Constraints on the Intergalactic Magnetic Field from Gamma-Ray
  Observations of Blazars}}, {\emph{eConf} {\bfseries C121028} (2012) 365}
  [\href{https://arxiv.org/abs/1303.5093}{{\ttfamily 1303.5093}}].

\bibitem{Finke:2015ona}
J.~D. Finke, L.~C. Reyes, M.~Georganopoulos, K.~Reynolds, M.~Ajello, S.~J.
  Fegan and K.~McCann, \emph{{Constraints on the Intergalactic Magnetic Field
  with Gamma-Ray Observations of Blazars}},
  \href{https://doi.org/10.1088/0004-637X/814/1/20}{\emph{Astrophys. J.}
  {\bfseries 814} (2015) 20}
  [\href{https://arxiv.org/abs/1510.02485}{{\ttfamily 1510.02485}}].

\bibitem{Kandus:2010nw}
A.~Kandus, K.~E. Kunze and C.~G. Tsagas, \emph{{Primordial magnetogenesis}},
  \href{https://doi.org/10.1016/j.physrep.2011.03.001}{\emph{Phys. Rept.}
  {\bfseries 505} (2011) 1} [\href{https://arxiv.org/abs/1007.3891}{{\ttfamily
  1007.3891}}].

\bibitem{Widrow:2011hs}
L.~M. Widrow, D.~Ryu, D.~R.~G. Schleicher, K.~Subramanian, C.~G. Tsagas and
  R.~A. Treumann, \emph{{The First Magnetic Fields}},
  \href{https://doi.org/10.1007/s11214-011-9833-5}{\emph{Space Sci. Rev.}
  {\bfseries 166} (2012) 37} [\href{https://arxiv.org/abs/1109.4052}{{\ttfamily
  1109.4052}}].

\bibitem{Durrer:2013pga}
R.~Durrer and A.~Neronov, \emph{{Cosmological Magnetic Fields: Their
  Generation, Evolution and Observation}},
  \href{https://doi.org/10.1007/s00159-013-0062-7}{\emph{Astron. Astrophys.
  Rev.} {\bfseries 21} (2013) 62}
  [\href{https://arxiv.org/abs/1303.7121}{{\ttfamily 1303.7121}}].

\bibitem{Subramanian:2015lua}
K.~Subramanian, \emph{{The origin, evolution and signatures of primordial
  magnetic fields}},
  \href{https://doi.org/10.1088/0034-4885/79/7/076901}{\emph{Rept. Prog. Phys.}
  {\bfseries 79} (2016) 076901}
  [\href{https://arxiv.org/abs/1504.02311}{{\ttfamily 1504.02311}}].

\bibitem{Vachaspati:2020blt}
T.~Vachaspati, \emph{{Progress on cosmological magnetic fields}},
  \href{https://doi.org/10.1088/1361-6633/ac03a9}{\emph{Rept. Prog. Phys.}
  {\bfseries 84} (2021) 074901}
  [\href{https://arxiv.org/abs/2010.10525}{{\ttfamily 2010.10525}}].

\bibitem{Turner:1987bw}
M.~S. Turner and L.~M. Widrow, \emph{{Inflation Produced, Large Scale Magnetic
  Fields}}, \href{https://doi.org/10.1103/PhysRevD.37.2743}{\emph{Phys. Rev. D}
  {\bfseries 37} (1988) 2743}.

\bibitem{Ratra:1991bn}
B.~Ratra, \emph{{Cosmological 'seed' magnetic field from inflation}},
  \href{https://doi.org/10.1086/186384}{\emph{Astrophys. J. Lett.} {\bfseries
  391} (1992) L1}.

\bibitem{Carroll:1991zs}
S.~M. Carroll and G.~B. Field, \emph{{The Einstein equivalence principle and
  the polarization of radio galaxies}},
  \href{https://doi.org/10.1103/PhysRevD.43.3789}{\emph{Phys. Rev. D}
  {\bfseries 43} (1991) 3789}.

\bibitem{DiazGil:2007dy}
A.~Diaz-Gil, J.~Garcia-Bellido, M.~Garcia~Perez and A.~Gonzalez-Arroyo,
  \emph{{Magnetic field production during preheating at the electroweak
  scale}}, \href{https://doi.org/10.1103/PhysRevLett.100.241301}{\emph{Phys.
  Rev. Lett.} {\bfseries 100} (2008) 241301}
  [\href{https://arxiv.org/abs/0712.4263}{{\ttfamily 0712.4263}}].

\bibitem{Martin:2007ue}
J.~Martin and J.~Yokoyama, \emph{{Generation of Large-Scale Magnetic Fields in
  Single-Field Inflation}},
  \href{https://doi.org/10.1088/1475-7516/2008/01/025}{\emph{JCAP} {\bfseries
  01} (2008) 025} [\href{https://arxiv.org/abs/0711.4307}{{\ttfamily
  0711.4307}}].

\bibitem{Demozzi:2009fu}
V.~Demozzi, V.~Mukhanov and H.~Rubinstein, \emph{{Magnetic fields from
  inflation?}},
  \href{https://doi.org/10.1088/1475-7516/2009/08/025}{\emph{JCAP} {\bfseries
  08} (2009) 025} [\href{https://arxiv.org/abs/0907.1030}{{\ttfamily
  0907.1030}}].

\bibitem{Kanno:2009ei}
S.~Kanno, J.~Soda and M.-a. Watanabe, \emph{{Cosmological Magnetic Fields from
  Inflation and Backreaction}},
  \href{https://doi.org/10.1088/1475-7516/2009/12/009}{\emph{JCAP} {\bfseries
  12} (2009) 009} [\href{https://arxiv.org/abs/0908.3509}{{\ttfamily
  0908.3509}}].

\bibitem{Bamba:2006ga}
K.~Bamba and M.~Sasaki, \emph{{Large-scale magnetic fields in the inflationary
  universe}}, \href{https://doi.org/10.1088/1475-7516/2007/02/030}{\emph{JCAP}
  {\bfseries 02} (2007) 030}
  [\href{https://arxiv.org/abs/astro-ph/0611701}{{\ttfamily
  astro-ph/0611701}}].

\bibitem{Ng:2015ewp}
K.-W. Ng, S.-L. Cheng and W.~Lee, \emph{{Inflationary dilaton-axion
  magnetogenesis}}, \href{https://doi.org/10.6122/CJP.20150909}{\emph{Chin. J.
  Phys.} {\bfseries 53} (2015) 110105}
  [\href{https://arxiv.org/abs/1409.2656}{{\ttfamily 1409.2656}}].

\bibitem{Ferreira:2013sqa}
R.~J.~Z. Ferreira, R.~K. Jain and M.~S. Sloth, \emph{{Inflationary
  magnetogenesis without the strong coupling problem}},
  \href{https://doi.org/10.1088/1475-7516/2013/10/004}{\emph{JCAP} {\bfseries
  10} (2013) 004} [\href{https://arxiv.org/abs/1305.7151}{{\ttfamily
  1305.7151}}].

\bibitem{Ferreira:2014hma}
R.~J.~Z. Ferreira, R.~K. Jain and M.~S. Sloth, \emph{{Inflationary
  Magnetogenesis without the Strong Coupling Problem II: Constraints from CMB
  anisotropies and B-modes}},
  \href{https://doi.org/10.1088/1475-7516/2014/06/053}{\emph{JCAP} {\bfseries
  06} (2014) 053} [\href{https://arxiv.org/abs/1403.5516}{{\ttfamily
  1403.5516}}].

\bibitem{Campanelli:2013mea}
L.~Campanelli, \emph{{Origin of Cosmic Magnetic Fields}},
  \href{https://doi.org/10.1103/PhysRevLett.111.061301}{\emph{Phys. Rev. Lett.}
  {\bfseries 111} (2013) 061301}
  [\href{https://arxiv.org/abs/1304.6534}{{\ttfamily 1304.6534}}].

\bibitem{Adshead:2016iae}
P.~Adshead, J.~T. Giblin, T.~R. Scully and E.~I. Sfakianakis,
  \emph{{Magnetogenesis from axion inflation}},
  \href{https://doi.org/10.1088/1475-7516/2016/10/039}{\emph{JCAP} {\bfseries
  10} (2016) 039} [\href{https://arxiv.org/abs/1606.08474}{{\ttfamily
  1606.08474}}].

\bibitem{Sharma:2017eps}
R.~Sharma, S.~Jagannathan, T.~R. Seshadri and K.~Subramanian, \emph{{Challenges
  in Inflationary Magnetogenesis: Constraints from Strong Coupling,
  Backreaction and the Schwinger Effect}},
  \href{https://doi.org/10.1103/PhysRevD.96.083511}{\emph{Phys. Rev. D}
  {\bfseries 96} (2017) 083511}
  [\href{https://arxiv.org/abs/1708.08119}{{\ttfamily 1708.08119}}].

\bibitem{Fujita:2019pmi}
T.~Fujita and R.~Durrer, \emph{{Scale-invariant Helical Magnetic Fields from
  Inflation}}, \href{https://doi.org/10.1088/1475-7516/2019/09/008}{\emph{JCAP}
  {\bfseries 09} (2019) 008}
  [\href{https://arxiv.org/abs/1904.11428}{{\ttfamily 1904.11428}}].

\bibitem{Gorbar:2021zlr}
E.~V. Gorbar, K.~Schmitz, O.~O. Sobol and S.~I. Vilchinskii,
  \emph{{Hypermagnetogenesis from axion inflation: Model-independent
  estimates}}, \href{https://doi.org/10.1103/PhysRevD.105.043530}{\emph{Phys.
  Rev. D} {\bfseries 105} (2022) 043530}
  [\href{https://arxiv.org/abs/2111.04712}{{\ttfamily 2111.04712}}].

\bibitem{Kushwaha:2020nfa}
A.~Kushwaha and S.~Shankaranarayanan, \emph{{Helical magnetic fields from
  Riemann coupling}},
  \href{https://doi.org/10.1103/PhysRevD.102.103528}{\emph{Phys. Rev. D}
  {\bfseries 102} (2020) 103528}
  [\href{https://arxiv.org/abs/2008.10825}{{\ttfamily 2008.10825}}].

\bibitem{Maity:2021qps}
D.~Maity, S.~Pal and T.~Paul, \emph{{Effective Theory of Inflationary
  Magnetogenesis and Constraints on Reheating}},
  \href{https://doi.org/10.1088/1475-7516/2021/05/045}{\emph{JCAP} {\bfseries
  05} (2021) 045} [\href{https://arxiv.org/abs/2103.02411}{{\ttfamily
  2103.02411}}].

\bibitem{Tripathy:2021sfb}
S.~Tripathy, D.~Chowdhury, R.~K. Jain and L.~Sriramkumar, \emph{{Challenges in
  the choice of the nonconformal coupling function in inflationary
  magnetogenesis}},
  \href{https://doi.org/10.1103/PhysRevD.105.063519}{\emph{Phys. Rev. D}
  {\bfseries 105} (2022) 063519}
  [\href{https://arxiv.org/abs/2111.01478}{{\ttfamily 2111.01478}}].

\bibitem{Vachaspati:1991nm}
T.~Vachaspati, \emph{{Magnetic fields from cosmological phase transitions}},
  \href{https://doi.org/10.1016/0370-2693(91)90051-Q}{\emph{Phys. Lett. B}
  {\bfseries 265} (1991) 258}.

\bibitem{Sigl:1996dm}
G.~Sigl, A.~V. Olinto and K.~Jedamzik, \emph{{Primordial magnetic fields from
  cosmological first order phase transitions}},
  \href{https://doi.org/10.1103/PhysRevD.55.4582}{\emph{Phys. Rev. D}
  {\bfseries 55} (1997) 4582}
  [\href{https://arxiv.org/abs/astro-ph/9610201}{{\ttfamily
  astro-ph/9610201}}].

\bibitem{Dolgov:2001nv}
A.~D. Dolgov and D.~Grasso, \emph{{Generation of cosmic magnetic fields and
  gravitational waves at neutrino decoupling}},
  \href{https://doi.org/10.1103/PhysRevLett.88.011301}{\emph{Phys. Rev. Lett.}
  {\bfseries 88} (2002) 011301}
  [\href{https://arxiv.org/abs/astro-ph/0106154}{{\ttfamily
  astro-ph/0106154}}].

\bibitem{Stevens:2007ep}
T.~Stevens, M.~B. Johnson, L.~S. Kisslinger, E.~M. Henley, W.~Y.~P. Hwang and
  M.~Burkardt, \emph{{Role of Charged Gauge Fields in Generating Magnetic Seed
  Fields in Bubble Collisions during the Cosmological Electroweak Phase
  Transition}}, \href{https://doi.org/10.1103/PhysRevD.77.023501}{\emph{Phys.
  Rev. D} {\bfseries 77} (2008) 023501}
  [\href{https://arxiv.org/abs/0707.1346}{{\ttfamily 0707.1346}}].

\bibitem{Kahniashvili:2009qi}
T.~Kahniashvili, A.~G. Tevzadze and B.~Ratra, \emph{{Phase Transition Generated
  Cosmological Magnetic Field at Large Scales}},
  \href{https://doi.org/10.1088/0004-637X/726/2/78}{\emph{Astrophys. J.}
  {\bfseries 726} (2011) 78} [\href{https://arxiv.org/abs/0907.0197}{{\ttfamily
  0907.0197}}].

\bibitem{Henley:2010ba}
E.~M. Henley, M.~B. Johnson and L.~S. Kisslinger, \emph{{EWPT Nucleation With
  MSSM and Electromagnetic Field Creation}},
  \href{https://doi.org/10.1103/PhysRevD.81.085035}{\emph{Phys. Rev. D}
  {\bfseries 81} (2010) 085035}
  [\href{https://arxiv.org/abs/1001.2783}{{\ttfamily 1001.2783}}].

\bibitem{Grasso:2000wj}
D.~Grasso and H.~R. Rubinstein, \emph{{Magnetic fields in the early universe}},
  \href{https://doi.org/10.1016/S0370-1573(00)00110-1}{\emph{Phys. Rept.}
  {\bfseries 348} (2001) 163}
  [\href{https://arxiv.org/abs/astro-ph/0009061}{{\ttfamily
  astro-ph/0009061}}].

\bibitem{Cheng:1994yr}
B.-l. Cheng and A.~V. Olinto, \emph{{Primordial magnetic fields generated in
  the quark - hadron transition}},
  \href{https://doi.org/10.1103/PhysRevD.50.2421}{\emph{Phys. Rev. D}
  {\bfseries 50} (1994) 2421}.

\bibitem{Balaji:2024rvo}
S.~Balaji, M.~Fairbairn and M.~O. Olea-Romacho, \emph{{Magnetogenesis with
  gravitational waves and primordial black hole dark matter}},
  \href{https://doi.org/10.1103/PhysRevD.109.075048}{\emph{Phys. Rev. D}
  {\bfseries 109} (2024) 075048}
  [\href{https://arxiv.org/abs/2402.05179}{{\ttfamily 2402.05179}}].

\bibitem{Barnaby:2012tk}
N.~Barnaby, R.~Namba and M.~Peloso, \emph{{Observable non-gaussianity from
  gauge field production in slow roll inflation, and a challenging connection
  with magnetogenesis}},
  \href{https://doi.org/10.1103/PhysRevD.85.123523}{\emph{Phys. Rev. D}
  {\bfseries 85} (2012) 123523}
  [\href{https://arxiv.org/abs/1202.1469}{{\ttfamily 1202.1469}}].

\bibitem{Campanelli:2008kh}
L.~Campanelli, \emph{{Helical Magnetic Fields from Inflation}},
  \href{https://doi.org/10.1142/S0218271809015175}{\emph{Int. J. Mod. Phys. D}
  {\bfseries 18} (2009) 1395}
  [\href{https://arxiv.org/abs/0805.0575}{{\ttfamily 0805.0575}}].

\bibitem{Campanelli:2007tc}
L.~Campanelli, \emph{{Evolution of Magnetic Fields in Freely Decaying
  Magnetohydrodynamic Turbulence}},
  \href{https://doi.org/10.1103/PhysRevLett.98.251302}{\emph{Phys. Rev. Lett.}
  {\bfseries 98} (2007) 251302}
  [\href{https://arxiv.org/abs/0705.2308}{{\ttfamily 0705.2308}}].

\bibitem{Banerjee:2004df}
R.~Banerjee and K.~Jedamzik, \emph{{The Evolution of cosmic magnetic fields:
  From the very early universe, to recombination, to the present}},
  \href{https://doi.org/10.1103/PhysRevD.70.123003}{\emph{Phys. Rev. D}
  {\bfseries 70} (2004) 123003}
  [\href{https://arxiv.org/abs/astro-ph/0410032}{{\ttfamily
  astro-ph/0410032}}].

\bibitem{Son:1998my}
D.~T. Son, \emph{{Magnetohydrodynamics of the early universe and the evolution
  of primordial magnetic fields}},
  \href{https://doi.org/10.1103/PhysRevD.59.063008}{\emph{Phys. Rev. D}
  {\bfseries 59} (1999) 063008}
  [\href{https://arxiv.org/abs/hep-ph/9803412}{{\ttfamily hep-ph/9803412}}].

\bibitem{Garretson:1992vt}
W.~D. Garretson, G.~B. Field and S.~M. Carroll, \emph{{Primordial magnetic
  fields from pseudoGoldstone bosons}},
  \href{https://doi.org/10.1103/PhysRevD.46.5346}{\emph{Phys. Rev. D}
  {\bfseries 46} (1992) 5346}
  [\href{https://arxiv.org/abs/hep-ph/9209238}{{\ttfamily hep-ph/9209238}}].

\bibitem{Ferreira:2014zia}
R.~Z. Ferreira and M.~S. Sloth, \emph{{Universal Constraints on Axions from
  Inflation}}, \href{https://doi.org/10.1007/JHEP12(2014)139}{\emph{JHEP}
  {\bfseries 12} (2014) 139} [\href{https://arxiv.org/abs/1409.5799}{{\ttfamily
  1409.5799}}].

\bibitem{Prokopec:2001nc}
T.~Prokopec, \emph{{Cosmological magnetic fields from photon coupling to
  fermions and bosons in inflation}},
  \href{https://arxiv.org/abs/astro-ph/0106247}{{\ttfamily astro-ph/0106247}}.

\bibitem{Anber:2009ua}
M.~M. Anber and L.~Sorbo, \emph{{Naturally inflating on steep potentials
  through electromagnetic dissipation}},
  \href{https://doi.org/10.1103/PhysRevD.81.043534}{\emph{Phys. Rev. D}
  {\bfseries 81} (2010) 043534}
  [\href{https://arxiv.org/abs/0908.4089}{{\ttfamily 0908.4089}}].

\bibitem{Barnaby:2011vw}
N.~Barnaby, R.~Namba and M.~Peloso, \emph{{Phenomenology of a Pseudo-Scalar
  Inflaton: Naturally Large Nongaussianity}},
  \href{https://doi.org/10.1088/1475-7516/2011/04/009}{\emph{JCAP} {\bfseries
  04} (2011) 009} [\href{https://arxiv.org/abs/1102.4333}{{\ttfamily
  1102.4333}}].

\bibitem{Barnaby:2011qe}
N.~Barnaby, E.~Pajer and M.~Peloso, \emph{{Gauge Field Production in Axion
  Inflation: Consequences for Monodromy, non-Gaussianity in the CMB, and
  Gravitational Waves at Interferometers}},
  \href{https://doi.org/10.1103/PhysRevD.85.023525}{\emph{Phys. Rev. D}
  {\bfseries 85} (2012) 023525}
  [\href{https://arxiv.org/abs/1110.3327}{{\ttfamily 1110.3327}}].

\bibitem{Shiraishi:2013kxa}
M.~Shiraishi, A.~Ricciardone and S.~Saga, \emph{{Parity violation in the CMB
  bispectrum by a rolling pseudoscalar}},
  \href{https://doi.org/10.1088/1475-7516/2013/11/051}{\emph{JCAP} {\bfseries
  11} (2013) 051} [\href{https://arxiv.org/abs/1308.6769}{{\ttfamily
  1308.6769}}].

\bibitem{Cook:2013xea}
J.~L. Cook and L.~Sorbo, \emph{{An inflationary model with small scalar and
  large tensor nongaussianities}},
  \href{https://doi.org/10.1088/1475-7516/2013/11/047}{\emph{JCAP} {\bfseries
  11} (2013) 047} [\href{https://arxiv.org/abs/1307.7077}{{\ttfamily
  1307.7077}}].

\bibitem{Ferreira:2015omg}
R.~Z. Ferreira, J.~Ganc, J.~Nore\~na and M.~S. Sloth, \emph{{On the validity of
  the perturbative description of axions during inflation}},
  \href{https://doi.org/10.1088/1475-7516/2016/04/039}{\emph{JCAP} {\bfseries
  04} (2016) 039} [\href{https://arxiv.org/abs/1512.06116}{{\ttfamily
  1512.06116}}]. [Erratum: JCAP 10, E01 (2016)].

\bibitem{Peloso:2016gqs}
M.~Peloso, L.~Sorbo and C.~Unal, \emph{{Rolling axions during inflation:
  perturbativity and signatures}},
  \href{https://doi.org/10.1088/1475-7516/2016/09/001}{\emph{JCAP} {\bfseries
  09} (2016) 001} [\href{https://arxiv.org/abs/1606.00459}{{\ttfamily
  1606.00459}}].

\bibitem{Durrer:2024ibi}
R.~Durrer, R.~von Eckardstein, D.~Garg, K.~Schmitz, O.~Sobol and
  S.~Vilchinskii, \emph{{Scalar perturbations from inflation in the presence of
  gauge fields}},  \href{https://arxiv.org/abs/2404.19694}{{\ttfamily
  2404.19694}}.

\bibitem{Cheng:2015oqa}
S.-L. Cheng, W.~Lee and K.-W. Ng, \emph{{Numerical study of pseudoscalar
  inflation with an axion-gauge field coupling}},
  \href{https://doi.org/10.1103/PhysRevD.93.063510}{\emph{Phys. Rev. D}
  {\bfseries 93} (2016) 063510}
  [\href{https://arxiv.org/abs/1508.00251}{{\ttfamily 1508.00251}}].

\bibitem{Notari:2016npn}
A.~Notari and K.~Tywoniuk, \emph{{Dissipative Axial Inflation}},
  \href{https://doi.org/10.1088/1475-7516/2016/12/038}{\emph{JCAP} {\bfseries
  12} (2016) 038} [\href{https://arxiv.org/abs/1608.06223}{{\ttfamily
  1608.06223}}].

\bibitem{Sobol:2019xls}
O.~O. Sobol, E.~V. Gorbar and S.~I. Vilchinskii, \emph{{Backreaction of
  electromagnetic fields and the Schwinger effect in pseudoscalar inflation
  magnetogenesis}},
  \href{https://doi.org/10.1103/PhysRevD.100.063523}{\emph{Phys. Rev. D}
  {\bfseries 100} (2019) 063523}
  [\href{https://arxiv.org/abs/1907.10443}{{\ttfamily 1907.10443}}].

\bibitem{DallAgata:2019yrr}
G.~Dall'Agata, S.~Gonz\'alez-Mart\'\i{}n, A.~Papageorgiou and M.~Peloso,
  \emph{{Warm dark energy}},
  \href{https://doi.org/10.1088/1475-7516/2020/08/032}{\emph{JCAP} {\bfseries
  08} (2020) 032} [\href{https://arxiv.org/abs/1912.09950}{{\ttfamily
  1912.09950}}].

\bibitem{Domcke:2020zez}
V.~Domcke, V.~Guidetti, Y.~Welling and A.~Westphal, \emph{{Resonant
  backreaction in axion inflation}},
  \href{https://doi.org/10.1088/1475-7516/2020/09/009}{\emph{JCAP} {\bfseries
  09} (2020) 009} [\href{https://arxiv.org/abs/2002.02952}{{\ttfamily
  2002.02952}}].

\bibitem{Peloso:2022ovc}
M.~Peloso and L.~Sorbo, \emph{{Instability in axion inflation with strong
  backreaction from gauge modes}},
  \href{https://doi.org/10.1088/1475-7516/2023/01/038}{\emph{JCAP} {\bfseries
  01} (2023) 038} [\href{https://arxiv.org/abs/2209.08131}{{\ttfamily
  2209.08131}}].

\bibitem{Caravano:2022epk}
A.~Caravano, E.~Komatsu, K.~D. Lozanov and J.~Weller, \emph{{Lattice
  simulations of axion-U(1) inflation}},
  \href{https://doi.org/10.1103/PhysRevD.108.043504}{\emph{Phys. Rev. D}
  {\bfseries 108} (2023) 043504}
  [\href{https://arxiv.org/abs/2204.12874}{{\ttfamily 2204.12874}}].

\bibitem{Garcia-Bellido:2023ser}
J.~Garcia-Bellido, A.~Papageorgiou, M.~Peloso and L.~Sorbo, \emph{{A flashing
  beacon in axion inflation: recurring bursts of gravitational waves in the
  strong backreaction regime}},
  \href{https://arxiv.org/abs/2303.13425}{{\ttfamily 2303.13425}}.

\bibitem{vonEckardstein:2023gwk}
R.~von Eckardstein, M.~Peloso, K.~Schmitz, O.~Sobol and L.~Sorbo, \emph{{Axion
  inflation in the strong-backreaction regime: decay of the Anber-Sorbo
  solution}}, \href{https://doi.org/10.1007/JHEP11(2023)183}{\emph{JHEP}
  {\bfseries 11} (2023) 183}
  [\href{https://arxiv.org/abs/2309.04254}{{\ttfamily 2309.04254}}].

\bibitem{Figueroa:2023oxc}
D.~G. Figueroa, J.~Lizarraga, A.~Urio and J.~Urrestilla, \emph{{Strong
  Backreaction Regime in Axion Inflation}},
  \href{https://doi.org/10.1103/PhysRevLett.131.151003}{\emph{Phys. Rev. Lett.}
  {\bfseries 131} (2023) 151003}
  [\href{https://arxiv.org/abs/2303.17436}{{\ttfamily 2303.17436}}].

\bibitem{Sharma:2024nfu}
R.~Sharma, A.~Brandenburg, K.~Subramanian and A.~Vikman, \emph{{Lattice
  simulations of axion-U(1) inflation: gravitational waves, magnetic fields,
  and black holes}},  \href{https://arxiv.org/abs/2411.04854}{{\ttfamily
  2411.04854}}.

\bibitem{Barnaby:2010vf}
N.~Barnaby and M.~Peloso, \emph{{Large Nongaussianity in Axion Inflation}},
  \href{https://doi.org/10.1103/PhysRevLett.106.181301}{\emph{Phys. Rev. Lett.}
  {\bfseries 106} (2011) 181301}
  [\href{https://arxiv.org/abs/1011.1500}{{\ttfamily 1011.1500}}].

\bibitem{Caravano:2024xsb}
A.~Caravano and M.~Peloso, \emph{{Unveiling the nonlinear dynamics of a rolling
  axion during inflation}},  \href{https://arxiv.org/abs/2407.13405}{{\ttfamily
  2407.13405}}.

\bibitem{Linde:2012bt}
A.~Linde, S.~Mooij and E.~Pajer, \emph{{Gauge field production in supergravity
  inflation: Local non-Gaussianity and primordial black holes}},
  \href{https://doi.org/10.1103/PhysRevD.87.103506}{\emph{Phys. Rev. D}
  {\bfseries 87} (2013) 103506}
  [\href{https://arxiv.org/abs/1212.1693}{{\ttfamily 1212.1693}}].

\bibitem{Bugaev:2013fya}
E.~Bugaev and P.~Klimai, \emph{{Axion inflation with gauge field production and
  primordial black holes}},
  \href{https://doi.org/10.1103/PhysRevD.90.103501}{\emph{Phys. Rev. D}
  {\bfseries 90} (2014) 103501}
  [\href{https://arxiv.org/abs/1312.7435}{{\ttfamily 1312.7435}}].

\bibitem{Freese:1990rb}
K.~Freese, J.~A. Frieman and A.~V. Olinto, \emph{{Natural inflation with pseudo
  - Nambu-Goldstone bosons}},
  \href{https://doi.org/10.1103/PhysRevLett.65.3233}{\emph{Phys. Rev. Lett.}
  {\bfseries 65} (1990) 3233}.

\bibitem{Maleknejad:2011jw}
A.~Maleknejad and M.~M. Sheikh-Jabbari, \emph{{Gauge-flation: Inflation From
  Non-Abelian Gauge Fields}},
  \href{https://doi.org/10.1016/j.physletb.2013.05.001}{\emph{Phys. Lett. B}
  {\bfseries 723} (2013) 224}
  [\href{https://arxiv.org/abs/1102.1513}{{\ttfamily 1102.1513}}].

\bibitem{Maleknejad:2011sq}
A.~Maleknejad and M.~M. Sheikh-Jabbari, \emph{{Non-Abelian Gauge Field
  Inflation}}, \href{https://doi.org/10.1103/PhysRevD.84.043515}{\emph{Phys.
  Rev. D} {\bfseries 84} (2011) 043515}
  [\href{https://arxiv.org/abs/1102.1932}{{\ttfamily 1102.1932}}].

\bibitem{Adshead:2013nka}
P.~Adshead, E.~Martinec and M.~Wyman, \emph{{Perturbations in Chromo-Natural
  Inflation}}, \href{https://doi.org/10.1007/JHEP09(2013)087}{\emph{JHEP}
  {\bfseries 09} (2013) 087} [\href{https://arxiv.org/abs/1305.2930}{{\ttfamily
  1305.2930}}].

\bibitem{Adshead:2013qp}
P.~Adshead, E.~Martinec and M.~Wyman, \emph{{Gauge fields and inflation: Chiral
  gravitational waves, fluctuations, and the Lyth bound}},
  \href{https://doi.org/10.1103/PhysRevD.88.021302}{\emph{Phys. Rev. D}
  {\bfseries 88} (2013) 021302}
  [\href{https://arxiv.org/abs/1301.2598}{{\ttfamily 1301.2598}}].

\bibitem{Adshead:2012kp}
P.~Adshead and M.~Wyman, \emph{{Chromo-Natural Inflation: Natural inflation on
  a steep potential with classical non-Abelian gauge fields}},
  \href{https://doi.org/10.1103/PhysRevLett.108.261302}{\emph{Phys. Rev. Lett.}
  {\bfseries 108} (2012) 261302}
  [\href{https://arxiv.org/abs/1202.2366}{{\ttfamily 1202.2366}}].

\bibitem{Maleknejad:2016qjz}
A.~Maleknejad, \emph{{Axion Inflation with an SU(2) Gauge Field: Detectable
  Chiral Gravity Waves}},
  \href{https://doi.org/10.1007/JHEP07(2016)104}{\emph{JHEP} {\bfseries 07}
  (2016) 104} [\href{https://arxiv.org/abs/1604.03327}{{\ttfamily
  1604.03327}}].

\bibitem{Papageorgiou:2018rfx}
A.~Papageorgiou, M.~Peloso and C.~Unal, \emph{{Nonlinear perturbations from the
  coupling of the inflaton to a non-Abelian gauge field, with a focus on
  Chromo-Natural Inflation}},
  \href{https://doi.org/10.1088/1475-7516/2018/09/030}{\emph{JCAP} {\bfseries
  09} (2018) 030} [\href{https://arxiv.org/abs/1806.08313}{{\ttfamily
  1806.08313}}].

\bibitem{Adshead:2016omu}
P.~Adshead, E.~Martinec, E.~I. Sfakianakis and M.~Wyman, \emph{{Higgsed
  Chromo-Natural Inflation}},
  \href{https://doi.org/10.1007/JHEP12(2016)137}{\emph{JHEP} {\bfseries 12}
  (2016) 137} [\href{https://arxiv.org/abs/1609.04025}{{\ttfamily
  1609.04025}}].

\bibitem{Lyth:1996im}
D.~H. Lyth, \emph{{What would we learn by detecting a gravitational wave signal
  in the cosmic microwave background anisotropy?}},
  \href{https://doi.org/10.1103/PhysRevLett.78.1861}{\emph{Phys. Rev. Lett.}
  {\bfseries 78} (1997) 1861}
  [\href{https://arxiv.org/abs/hep-ph/9606387}{{\ttfamily hep-ph/9606387}}].

\bibitem{Caldwell:2017chz}
R.~R. Caldwell and C.~Devulder, \emph{{Axion Gauge Field Inflation and
  Gravitational Leptogenesis: A Lower Bound on B Modes from the
  Matter-Antimatter Asymmetry of the Universe}},
  \href{https://doi.org/10.1103/PhysRevD.97.023532}{\emph{Phys. Rev. D}
  {\bfseries 97} (2018) 023532}
  [\href{https://arxiv.org/abs/1706.03765}{{\ttfamily 1706.03765}}].

\bibitem{Obata:2014loa}
I.~Obata, T.~Miura and J.~Soda, \emph{{Chromo-Natural Inflation in the
  Axiverse}}, \href{https://doi.org/10.1103/PhysRevD.92.063516}{\emph{Phys.
  Rev. D} {\bfseries 92} (2015) 063516}
  [\href{https://arxiv.org/abs/1412.7620}{{\ttfamily 1412.7620}}]. [Addendum:
  Phys.Rev.D 95, 109902 (2017)].

\bibitem{Obata:2016tmo}
I.~Obata and J.~Soda, \emph{{Chiral primordial Chiral primordial gravitational
  waves from dilaton induced delayed chromonatural inflation}},
  \href{https://doi.org/10.1103/PhysRevD.93.123502}{\emph{Phys. Rev. D}
  {\bfseries 93} (2016) 123502}
  [\href{https://arxiv.org/abs/1602.06024}{{\ttfamily 1602.06024}}]. [Addendum:
  Phys.Rev.D 95, 109903 (2017)].

\bibitem{Domcke:2018rvv}
V.~Domcke, B.~Mares, F.~Muia and M.~Pieroni, \emph{{Emerging chromo-natural
  inflation}}, \href{https://doi.org/10.1088/1475-7516/2019/04/034}{\emph{JCAP}
  {\bfseries 04} (2019) 034}
  [\href{https://arxiv.org/abs/1807.03358}{{\ttfamily 1807.03358}}].

\bibitem{Dimastrogiovanni:2023oid}
E.~Dimastrogiovanni, M.~Fasiello, M.~Michelotti and L.~Pinol, \emph{{Primordial
  gravitational waves in non-minimally coupled chromo-natural inflation}},
  \href{https://doi.org/10.1088/1475-7516/2024/02/039}{\emph{JCAP} {\bfseries
  02} (2024) 039} [\href{https://arxiv.org/abs/2303.10718}{{\ttfamily
  2303.10718}}].

\bibitem{Murata:2024urv}
T.~Murata and T.~Kobayashi, \emph{{Chromo-natural inflation supported by
  enhanced friction from Horndeski gravity}},
  \href{https://arxiv.org/abs/2408.01773}{{\ttfamily 2408.01773}}.

\bibitem{Adshead:2012qe}
P.~Adshead and M.~Wyman, \emph{{Gauge-flation trajectories in Chromo-Natural
  Inflation}}, \href{https://doi.org/10.1103/PhysRevD.86.043530}{\emph{Phys.
  Rev. D} {\bfseries 86} (2012) 043530}
  [\href{https://arxiv.org/abs/1203.2264}{{\ttfamily 1203.2264}}].

\bibitem{Sheikh-Jabbari:2012tom}
M.~M. Sheikh-Jabbari, \emph{{Gauge-flation Vs Chromo-Natural Inflation}},
  \href{https://doi.org/10.1016/j.physletb.2012.09.014}{\emph{Phys. Lett. B}
  {\bfseries 717} (2012) 6} [\href{https://arxiv.org/abs/1203.2265}{{\ttfamily
  1203.2265}}].

\bibitem{Maleknejad:2012dt}
A.~Maleknejad and M.~Zarei, \emph{{Slow-roll trajectories in Chromo-Natural and
  Gauge-flation Models, an exhaustive analysis}},
  \href{https://doi.org/10.1103/PhysRevD.88.043509}{\emph{Phys. Rev. D}
  {\bfseries 88} (2013) 043509}
  [\href{https://arxiv.org/abs/1212.6760}{{\ttfamily 1212.6760}}].

\bibitem{Maleknejad:2012fw}
A.~Maleknejad, M.~M. Sheikh-Jabbari and J.~Soda, \emph{{Gauge Fields and
  Inflation}}, \href{https://doi.org/10.1016/j.physrep.2013.03.003}{\emph{Phys.
  Rept.} {\bfseries 528} (2013) 161}
  [\href{https://arxiv.org/abs/1212.2921}{{\ttfamily 1212.2921}}].

\bibitem{Iarygina:2021bxq}
O.~Iarygina and E.~I. Sfakianakis, \emph{{Gravitational waves from spectator
  Gauge-flation}},
  \href{https://doi.org/10.1088/1475-7516/2021/11/023}{\emph{JCAP} {\bfseries
  11} (2021) 023} [\href{https://arxiv.org/abs/2105.06972}{{\ttfamily
  2105.06972}}].

\bibitem{Dimastrogiovanni:2016fuu}
E.~Dimastrogiovanni, M.~Fasiello and T.~Fujita, \emph{{Primordial Gravitational
  Waves from Axion-Gauge Fields Dynamics}},
  \href{https://doi.org/10.1088/1475-7516/2017/01/019}{\emph{JCAP} {\bfseries
  01} (2017) 019} [\href{https://arxiv.org/abs/1608.04216}{{\ttfamily
  1608.04216}}].

\bibitem{Fujita:2017jwq}
T.~Fujita, R.~Namba and Y.~Tada, \emph{{Does the detection of primordial
  gravitational waves exclude low energy inflation?}},
  \href{https://doi.org/10.1016/j.physletb.2017.12.014}{\emph{Phys. Lett. B}
  {\bfseries 778} (2018) 17}
  [\href{https://arxiv.org/abs/1705.01533}{{\ttfamily 1705.01533}}].

\bibitem{Fujita:2018ndp}
T.~Fujita, E.~I. Sfakianakis and M.~Shiraishi, \emph{{Tensor Spectra Templates
  for Axion-Gauge Fields Dynamics during Inflation}},
  \href{https://doi.org/10.1088/1475-7516/2019/05/057}{\emph{JCAP} {\bfseries
  05} (2019) 057} [\href{https://arxiv.org/abs/1812.03667}{{\ttfamily
  1812.03667}}].

\bibitem{Iarygina:2023mtj}
O.~Iarygina, E.~I. Sfakianakis, R.~Sharma and A.~Brandenburg,
  \emph{{Backreaction of axion-SU(2) dynamics during inflation}},
  \href{https://doi.org/10.1088/1475-7516/2024/04/018}{\emph{JCAP} {\bfseries
  04} (2024) 018} [\href{https://arxiv.org/abs/2311.07557}{{\ttfamily
  2311.07557}}].

\bibitem{Dimastrogiovanni:2024xvc}
E.~Dimastrogiovanni, M.~Fasiello and A.~Papageorgiou, \emph{{A novel PBH
  production mechanism from non-Abelian gauge fields during inflation}},
  \href{https://arxiv.org/abs/2403.13581}{{\ttfamily 2403.13581}}.

\bibitem{Dimastrogiovanni:2024lzj}
E.~Dimastrogiovanni, M.~Fasiello, M.~Michelotti and O.~\"Ozsoy, \emph{{A
  universal constraint on axion non-Abelian dynamics during inflation}},
  \href{https://arxiv.org/abs/2405.17411}{{\ttfamily 2405.17411}}.

\bibitem{Alexander:2023flr}
S.~Alexander, C.~Creque-Sarbinowski, H.~Gilmer and K.~Freese, \emph{{Higgs
  Inflation and the Electroweak Gauge Sector}},
  \href{https://arxiv.org/abs/2306.04671}{{\ttfamily 2306.04671}}.

\bibitem{Brandenburg:2017neh}
A.~Brandenburg, T.~Kahniashvili, S.~Mandal, A.~Roper~Pol, A.~G. Tevzadze and
  T.~Vachaspati, \emph{{Evolution of hydromagnetic turbulence from the
  electroweak phase transition}},
  \href{https://doi.org/10.1103/PhysRevD.96.123528}{\emph{Phys. Rev. D}
  {\bfseries 96} (2017) 123528}
  [\href{https://arxiv.org/abs/1711.03804}{{\ttfamily 1711.03804}}].

\bibitem{Kamada:2020bmb}
K.~Kamada, F.~Uchida and J.~Yokoyama, \emph{{Baryon isocurvature constraints on
  the primordial hypermagnetic fields}},
  \href{https://doi.org/10.1088/1475-7516/2021/04/034}{\emph{JCAP} {\bfseries
  04} (2021) 034} [\href{https://arxiv.org/abs/2012.14435}{{\ttfamily
  2012.14435}}].

\bibitem{Kamada:2016cnb}
K.~Kamada and A.~J. Long, \emph{{Evolution of the Baryon Asymmetry through the
  Electroweak Crossover in the Presence of a Helical Magnetic Field}},
  \href{https://doi.org/10.1103/PhysRevD.94.123509}{\emph{Phys. Rev. D}
  {\bfseries 94} (2016) 123509}
  [\href{https://arxiv.org/abs/1610.03074}{{\ttfamily 1610.03074}}].

\bibitem{Maleknejad:2013npa}
A.~Maleknejad and E.~Erfani, \emph{{Chromo-Natural Model in Anisotropic
  Background}},
  \href{https://doi.org/10.1088/1475-7516/2014/03/016}{\emph{JCAP} {\bfseries
  03} (2014) 016} [\href{https://arxiv.org/abs/1311.3361}{{\ttfamily
  1311.3361}}].

\bibitem{Papageorgiou:2019ecb}
A.~Papageorgiou, M.~Peloso and C.~Unal, \emph{{Nonlinear perturbations from
  axion-gauge fields dynamics during inflation}},
  \href{https://doi.org/10.1088/1475-7516/2019/07/004}{\emph{JCAP} {\bfseries
  07} (2019) 004} [\href{https://arxiv.org/abs/1904.01488}{{\ttfamily
  1904.01488}}].

\bibitem{Dimastrogiovanni:2012ew}
E.~Dimastrogiovanni and M.~Peloso, \emph{{Stability analysis of chromo-natural
  inflation and possible evasion of Lyth\textquoteright{}s bound}},
  \href{https://doi.org/10.1103/PhysRevD.87.103501}{\emph{Phys. Rev. D}
  {\bfseries 87} (2013) 103501}
  [\href{https://arxiv.org/abs/1212.5184}{{\ttfamily 1212.5184}}].

\bibitem{Namba:2015gja}
R.~Namba, M.~Peloso, M.~Shiraishi, L.~Sorbo and C.~Unal, \emph{{Scale-dependent
  gravitational waves from a rolling axion}},
  \href{https://doi.org/10.1088/1475-7516/2016/01/041}{\emph{JCAP} {\bfseries
  01} (2016) 041} [\href{https://arxiv.org/abs/1509.07521}{{\ttfamily
  1509.07521}}].

\bibitem{Maleknejad:2018nxz}
A.~Maleknejad and E.~Komatsu, \emph{{Production and Backreaction of Spin-2
  Particles of $SU(2)$ Gauge Field during Inflation}},
  \href{https://doi.org/10.1007/JHEP05(2019)174}{\emph{JHEP} {\bfseries 05}
  (2019) 174} [\href{https://arxiv.org/abs/1808.09076}{{\ttfamily
  1808.09076}}].

\bibitem{Ishiwata:2021yne}
K.~Ishiwata, E.~Komatsu and I.~Obata, \emph{{Axion-gauge field dynamics with
  backreaction}},
  \href{https://doi.org/10.1088/1475-7516/2022/03/010}{\emph{JCAP} {\bfseries
  03} (2022) 010} [\href{https://arxiv.org/abs/2111.14429}{{\ttfamily
  2111.14429}}].

\bibitem{Kallosh:2013hoa}
R.~Kallosh and A.~Linde, \emph{{Universality Class in Conformal Inflation}},
  \href{https://doi.org/10.1088/1475-7516/2013/07/002}{\emph{JCAP} {\bfseries
  07} (2013) 002} [\href{https://arxiv.org/abs/1306.5220}{{\ttfamily
  1306.5220}}].

\bibitem{Kallosh:2013yoa}
R.~Kallosh, A.~Linde and D.~Roest, \emph{{Superconformal Inflationary
  $\alpha$-Attractors}},
  \href{https://doi.org/10.1007/JHEP11(2013)198}{\emph{JHEP} {\bfseries 11}
  (2013) 198} [\href{https://arxiv.org/abs/1311.0472}{{\ttfamily 1311.0472}}].

\bibitem{JOSS}
{Pencil Code Collaboration}, A.~{Brandenburg}, A.~{Johansen}, P.~{Bourdin},
  W.~{Dobler}, W.~{Lyra}, M.~{Rheinhardt}, S.~{Bingert}, N.~{Haugen}, A.~{Mee},
  F.~{Gent}, N.~{Babkovskaia}, C.-C. {Yang}, T.~{Heinemann}, B.~{Dintrans},
  D.~{Mitra}, S.~{Candelaresi}, J.~{Warnecke}, P.~{K{\"a}pyl{\"a}},
  A.~{Schreiber}, P.~{Chatterjee}, M.~{K{\"a}pyl{\"a}}, X.-Y. {Li},
  J.~{Kr{\"u}ger}, J.~{Aarnes}, G.~{Sarson}, J.~{Oishi}, J.~{Schober},
  R.~{Plasson}, C.~{Sandin}, E.~{Karchniwy}, L.~{Rodrigues}, A.~{Hubbard},
  G.~{Guerrero}, A.~{Snodin}, I.~{Losada}, J.~{Pekkil{\"a}} and C.~{Qian},
  \emph{{The Pencil Code, a modular MPI code for partial differential equations
  and particles: multipurpose and multiuser-maintained}},
  \href{https://doi.org/10.21105/joss.02807}{\emph{J. Open Source Software}
  {\bfseries 6} (2021) 2807}.

\bibitem{Adshead:2015pva}
P.~Adshead, J.~T. Giblin, T.~R. Scully and E.~I. Sfakianakis,
  \emph{{Gauge-preheating and the end of axion inflation}},
  \href{https://doi.org/10.1088/1475-7516/2015/12/034}{\emph{JCAP} {\bfseries
  12} (2015) 034} [\href{https://arxiv.org/abs/1502.06506}{{\ttfamily
  1502.06506}}].

\bibitem{Sharma:2018kgs}
R.~Sharma, K.~Subramanian and T.~R. Seshadri, \emph{{Generation of helical
  magnetic field in a viable scenario of inflationary magnetogenesis}},
  \href{https://doi.org/10.1103/PhysRevD.97.083503}{\emph{Phys. Rev. D}
  {\bfseries 97} (2018) 083503}
  [\href{https://arxiv.org/abs/1802.04847}{{\ttfamily 1802.04847}}].

\bibitem{MAGIC:2022piy}
{\scshape MAGIC} Collaboration, V.~A. Acciari et~al., \emph{{A lower bound on
  intergalactic magnetic fields from time variability of 1ES 0229+200 from
  MAGIC and Fermi/LAT observations}},
  \href{https://doi.org/10.1051/0004-6361/202244126}{\emph{Astron. Astrophys.}
  {\bfseries 670} (2023) A145}
  [\href{https://arxiv.org/abs/2210.03321}{{\ttfamily 2210.03321}}].

\bibitem{Planck:2015zrl}
{\scshape Planck} Collaboration, P.~A.~R. Ade et~al., \emph{{Planck 2015
  results. XIX. Constraints on primordial magnetic fields}},
  \href{https://doi.org/10.1051/0004-6361/201525821}{\emph{Astron. Astrophys.}
  {\bfseries 594} (2016) A19}
  [\href{https://arxiv.org/abs/1502.01594}{{\ttfamily 1502.01594}}].

\bibitem{Kronberg:1993vk}
P.~P. Kronberg, \emph{{Extragalactic magnetic fields}},
  \href{https://doi.org/10.1088/0034-4885/57/4/001}{\emph{Rept. Prog. Phys.}
  {\bfseries 57} (1994) 325}.

\bibitem{Fujita:2015iga}
T.~Fujita, R.~Namba, Y.~Tada, N.~Takeda and H.~Tashiro, \emph{{Consistent
  generation of magnetic fields in axion inflation models}},
  \href{https://doi.org/10.1088/1475-7516/2015/05/054}{\emph{JCAP} {\bfseries
  05} (2015) 054} [\href{https://arxiv.org/abs/1503.05802}{{\ttfamily
  1503.05802}}].

\bibitem{Gross:1980br}
D.~J. Gross, R.~D. Pisarski and L.~G. Yaffe, \emph{{QCD and Instantons at
  Finite Temperature}},
  \href{https://doi.org/10.1103/RevModPhys.53.43}{\emph{Rev. Mod. Phys.}
  {\bfseries 53} (1981) 43}.

\bibitem{Maleknejad:2020pec}
A.~Maleknejad, \emph{{Chiral anomaly in SU(2)$_{R}$-axion inflation and the new
  prediction for particle cosmology}},
  \href{https://doi.org/10.1007/JHEP06(2021)113}{\emph{JHEP} {\bfseries 21}
  (2020) 113} [\href{https://arxiv.org/abs/2103.14611}{{\ttfamily
  2103.14611}}].

\bibitem{Patel:2023ybi}
T.~Patel and T.~Vachaspati, \emph{{Annihilation of electroweak dumbbells}},
  \href{https://doi.org/10.1007/JHEP02(2024)164}{\emph{JHEP} {\bfseries 02}
  (2024) 164} [\href{https://arxiv.org/abs/2311.00026}{{\ttfamily
  2311.00026}}].

\end{thebibliography}\endgroup

\end{document}